\documentclass[showpacs,showkeys,nofootinbib]{revtex4}
\usepackage{graphicx,epsfig,wrapfig,color}
\newcommand{\be}{\begin{equation}}
\newcommand{\ee}{\end{equation}}
\newcommand{\ba}{\begin{eqnarray}}
\newcommand{\ea}{\end{eqnarray}}
\newcommand{\di}{\!{\rm d}}
\newcommand{\la}{\langle}
\newcommand{\ra}{\rangle}
\newcommand{\kringel}[1]{ \;\stackrel{\circ}{#1}}
\newcommand{\rotcorr}[1]{\mbox{\it\"{#1}}}
\newcommand{\fracS}[2]{{\textstyle\frac{#1}{#2}}}
\newcommand{\btau}{   {{\mbox{\boldmath$\tau$} }}}
\newcommand{\bDelta}{ {\bf\Delta}}
\begin{document}
\newcommand*{\Bochum}{
    Institut f\"ur Theoretische Physik II, Ruhr-Universit\"at Bochum,
    D-44780 Bochum, Germany}\affiliation{\Bochum}
\title{The nucleon form-factors of the energy momentum tensor in the Skyrme model}
\author{C.~Cebulla, K.~Goeke, J.~Ossmann, P.~Schweitzer}\affiliation{\Bochum }
\date{August, 2007}
\begin{abstract}
The nucleon form factors of the energy-momentum tensor are studied in the large-$N_c$
limit in the framework of the Skyrme model.
\end{abstract}



\maketitle
%
%
\section{Introduction}
\label{Sec-1:introduction}

The information contained in the nucleon form factors of the energy momentum
tensor (EMT) \cite{Pagels} allows to address, among others, questions like:
How are the total momentum and angular momentum  of the nucleon \cite{Ji:1996ek}
shared among its constituents?
Or, how are the strong forces experienced by its constituents distributed inside
the nucleon \cite{Polyakov:2002yz}?

An answer is known only to the first of these questions.
Deeply inelastic lepton nucleon scattering experiments show that
at scales of several ${\rm GeV}^2$ about half of the nucleon momentum 
(considered in the infinite-momentum frame)
is carried by quarks, the other half by gluons.
An appropriate tool allowing to deduce further information on the EMT form factors 
became available only during the last decade, namely generalized parton distribution 
functions \cite{GPDs} accessible in hard exclusive reactions
\cite{Saull:1999kt,Adloff:2001cn,Airapetian:2001yk,Stepanyan:2001sm,Ellinghaus:2002bq,Chekanov:2003ya,Aktas:2005ty,Airapetian:2006zr,Hall-A:2006hx},
see \cite{Ji:1998pc,Radyushkin:2000uy,Goeke:2001tz,Diehl:2003ny,Belitsky:2005qn}
for reviews.

The perspective to learn in this way about the EMT form factors
motivates theoretical studies.
The nucleon EMT form factors were studied in lattice QCD
\cite{Mathur:1999uf,Hagler:2003jd,Gockeler:2003jf,Negele:2004iu},
chiral perturbation theory \cite{Chen:2001pv,Belitsky:2002jp,Diehl:2006ya},
and models such as the chiral quark soliton model 
\cite{Wakamatsu:2006dy,Goeke:2007fp,Goeke:2007fq}.
(Particular aspects of the EMT form factors were discussed in this model in 
\cite{Petrov:1998kf,Kivel:2000fg,Schweitzer:2002nm,Ossmann:2004bp,Wakamatsu:2005vk}.)
In this work we study the nucleon EMT form factors in the Skyrme model
\cite{Skyrme:1961vq}.

The Skyrme model has important virtues. 
First, the model respects chiral symmetry which plays a major role in the
description of the nucleon. 
Second, it provides a practical realization of the picture of baryons as solitons 
of effective meson fields motivated in QCD in the limit of a large number of colours 
$N_c$ \cite{Witten:1979kh}. Though $N_c=3$ in nature does not seem to be a large 
number, large-$N_c$ relations are often found to agree with phenomenology,
although the approach has also its limitations \cite{large-Nc-reviews}.
Hereby it is also important to keep in mind that 
the chiral limit and the large-$N_c$ limit do not commute \cite{Dashen:1993jt}.
Last not least, the simplicity of the Skyrme model makes it an appealing ground 
for numerous theoretical studies 
\cite{Adkins:1983ya,Adkins:1983hy,Jackson:1983bi,Guadagnini:1983uv,Adkins:1984cf,Bander:1984gr,Braaten:1984qe,Braaten:1986iw,Ji:1991ff,Adkins:1983nw,Meissner:1987ge,Schwesinger:1988af},
see \cite{Zahed:1986qz,Holzwarth:1985rb,Meier:1996ng} for reviews.

The purpose of the present study is threefold. First, the study of the EMT
will shed some light on the model itself. Of particular interest in this
context is the stability of the Skyrmion. 
Second, we calculate the nucleon EMT form factors in the Skyrme model which are
of phenomenological interest for the studies of the hard exclusive reactions.
Third, we compare the Skyrme model results to the outcome from other chiral models,
in particular, from the chiral quark soliton model \cite{Goeke:2007fp,Goeke:2007fq}.
This comparison will help to recognize general features of the EMT form factors
which are common in the large-$N_c$ soliton approach and therefore likely to be
robust.

For completeness we remark that the general chiral structure of the EMT 
was discussed in chiral perturbation theory and/or chiral models in 
\cite{Donoghue:1991qv,Kubis:1999db,Megias:2004uj,Megias:2005fj}.
Issues of pion EMT form factors in lattice QCD were addressed in
\cite{Brommel:2005ee,Chen:2006gg}.

The note is organized as follows.
In Sec.~\ref{Sec-2:EMT-in-general} we introduce the nucleon EMT form factors.
In Sec.~\ref{Sec-3:model} we briefly review how the nucleon is described in 
the Skyrme model.
In Sec.~\ref{Sec-4:EMT-ffs-in-model} we derive the model expressions for the
form factors.
In Sec.~\ref{Sec-5:properties-of-the-densities-and-results} we discuss the 
densities of the static EMT, i.e.\  Fourier transforms of the form factors.
In \ref{Sec-6:ffs-chiral-properties} and \ref{Sec-7:Results-ffs} we discuss
the properties of the form factors and the numerical results,
and conclude in Sec.~\ref{Sec-8:conclusions}.
The Appendices~\ref{App:Alternative-definition}-\ref{App:scaling-relations}
contain remarks on the notation, technical details and proofs of the
theoretical consistency of the approach.

\section{Form factors of the energy-momentum tensor}
\label{Sec-2:EMT-in-general}

The nucleon matrix element of the total symmetric EMT are
characterized by three form factors as follows \cite{Pagels}
\be
    \la p^\prime| \hat T_{\mu\nu}(0) |p\rangle
    = \bar u(p^\prime)\biggl[M_2(t)\,\frac{P_\mu P_\nu}{M_N}+
    J(t)\ \frac{i(P_{\mu}\sigma_{\nu\rho}+P_{\nu}\sigma_{\mu\rho})
    \Delta^\rho}{2M_N}
    + d_1(t)\,
    \frac{\Delta_\mu\Delta_\nu-g_{\mu\nu}\Delta^2}{5M_N}\biggr]u(p)\, ,
    \label{Eq:ff-of-EMT} \ee
where $P=(p+p')/2$, $\Delta=(p'-p)$ and $t=\Delta^2$.
The normalizations
$\la p^\prime|p\ra = 2p^0(2\pi)^3\delta^{(3)}({\bf p}^\prime-{\bf p})$ and
$\bar u(p) u(p)=2 M_N$ are used, and spin indices are suppressed for brevity.

In QCD it is possible to define gauge invariant separate quark and gluon parts
of the EMT  having both decompositions analog to (\ref{Eq:ff-of-EMT})
with form factors $M_2^Q(t)$, $M_2^G(t)$, etc. In these decompositions,
however, in addition the terms appear $g_{\mu\nu}\bar c^{Q,G}(t)$ with
$\bar c^Q(t)=-\bar c^G(t)$ which account for non-conservation of the separate
quark and gluon parts of the EMT, as only the total EMT is conserved.
The quark (gluon) EMT form factors are related to Mellin moments
of unpolarized quark (gluon) generalized parton distribution functions
\cite{Ji:1996ek} which enter the description of certain hard exclusive reactions
\cite{Ji:1998pc,Radyushkin:2000uy,Goeke:2001tz,Diehl:2003ny,Belitsky:2005qn}.
In the following we shall focus on the total EMT.
For our approach it is of importance that in the large-$N_c$ limit the
nucleon EMT form factors behave as \cite{Goeke:2001tz}
\be\label{Eq:FF-large-Nc}
    M_2(t) = {\mathcal O}(N_c^0)\;,\;\;\;
    J(t)   = {\mathcal O}(N_c^0)\;,\;\;\;
    d_1(t) = {\mathcal O}(N_c^2)\;.
\ee
These relations hold also separately for the quark- and gluon-part of the EMT.

The form factors of the EMT in Eq.~(\ref{Eq:ff-of-EMT}) can be interpreted
\cite{Polyakov:2002yz} in analogy to the electromagnetic form factors
\cite{Sachs} in the Breit frame characterized by $\Delta^0=0$.
In this frame one can define the static EMT
\be\label{Def:static-EMT}
        T_{\mu\nu}({\bf r},{\bf s}) =
        \frac{1}{2p^0}
	\int\frac{\di^3\bDelta}{(2\pi)^3}\;\exp(i\bDelta{\bf r})\;
        \la p^\prime,S^\prime|\hat{T}_{\mu\nu}(0)|p,S\ra\;. \ee
The initial and final polarization vectors of the nucleon $S$ and $S^\prime$ are
defined such that in the respective rest-frame they are equal to $(0,{\bf s})$
with the unit vector ${\bf s}$ denoting the quantization axis for the spin.
The respective form factors are related to $T_{\mu\nu}({\bf r},{\bf s})$ by
(the following Eqs.~(\ref{Eq:ff-J},~\ref{Eq:ff-d1}) hold also separately
for quarks and gluons)
\ba
    J(t)+\frac{2t}{3}\, {J}^\prime(t)
    &=& \int\di^3{\bf r}\, e^{-i{\bf r}\bDelta}\,
    \varepsilon^{ijk}\,s_i\,r_j\,T_{0k}({\bf r},{\bf s})\, ,
    \label{Eq:ff-J}\\
    d_1(t)+\frac{4t}{3}\, d_1^{\,\prime}(t)
    +\frac{4t^2}{15}\, d_1^{\prime\prime}(t)
    &=& -\frac{M_N}{2}\, \int\di^3{\bf r}\,e^{-i{\bf r}\bDelta}\,
    T_{ij}({\bf r})\,\left(r^i r^j-\frac{{\bf r}^2}3\,\delta^{ij}\right)\,
    , \label{Eq:ff-d1}\\
    M_2(t)-\frac{t}{4M_N^2}\left(M_2(t)-2 J(t)+\frac 45\, d_1(t) \right)
    &=&\frac{1}{M_N}\,\int\di^3{\bf r}\, e^{-i{\bf r}\bDelta}
    \, T_{00}({\bf r},{\bf s})\,,
    \label{Eq:ff-M2}
\ea
where the primes denote derivatives with respect to $t$.
The form factors are renormalization scale independent and
satisfy at $t=0$ the constraints
\ba
    M_2(0)&=&
    \frac{1}{M_N}\,\int\di^3{\bf r}\;T_{00}({\bf r},{\bf s})=1
    \;,\nonumber\\
    J(0)&=&
    \int\di^3{\bf r}\;\varepsilon^{ijk}\,s_i\,r_j\,
    T_{0k}({\bf r},{\bf s})=\frac12\, , \nonumber\\
    d_1(0)&=&
    -\frac{M_N}{2}\, \int\di^3{\bf r}\;T_{ij}({\bf r})\,
    \left(r^i r^j-\frac{{\bf r}^2}3\,\delta^{ij}\right)\equiv d_1\,.
    \label{Eq:M2-J-d1}
\ea
The first two constraints mean that in the rest frame the total energy of the nucleon
is equal to its mass, and that its spin is 1/2. The value of $d_1$
is not known a priori and must be determined experimentally. However, being
a conserved quantity it is on equal footing with other characteristic nucleon
properties like mass, magnetic moment, etc. \cite{Polyakov:1999gs,Teryaev:2001qm}.

The components
$T_{00}({\bf r},{\bf s})$ and $\varepsilon^{ijk}r_jT_{0k}({\bf r},{\bf s})$
describe the distribution of nucleon momentum and angular momentum, while the
components $T_{ik}({\bf r})$ characterize the spatial distribution of strong
forces experienced by the partons inside the nucleon \cite{Polyakov:2002yz}.
In fact, $T_{ij}({\bf r})$ is the static stress tensor which
(for spin 0 and 1/2 particles) is given by
\be\label{Eq:T_ij-pressure-and-shear}
    T_{ij}({\bf r})
    = s(r)\left(\frac{r_ir_j}{r^2}-\frac 13\,\delta_{ij}\right)
        + p(r)\,\delta_{ij}\, . \ee
Here $p(r)$ describes the radial distribution of the pressure inside the
hadron, while $s(r)$ is related to the distribution of the shear forces
\cite{Polyakov:2002yz}. Both are related due to the conservation
of the EMT by the differential equation
\be\label{Eq:relation-p(r)-s(r)}
    \frac23\;\frac{\partial s(r)}{\partial r\;}+
    \frac{2s(r)}{r} + \frac{\partial p(r)}{\partial r\;} = 0\;.
\ee
Other important relations originating from conservation of EMT
are the stability condition
\be\label{Eq:stability}
    \int_0^\infty \!\di r\;r^2p(r)=0 \;,
\ee
and integral relations for $d_1$ in terms of $p(r)$ and $s(r)$
(see Appendix of Ref.~\cite{Goeke:2007fp} for further relations)
\be\label{Eq:d1-from-s(r)-and-p(r)}
        d_1  =  -\,\frac{1}{3}\;M_N \int\di^3{\bf r}\;r^2\, s(r)
         =     \frac{5}{4}\;M_N \int\di^3{\bf r}\;r^2\, p(r)\;.
\ee

It is worth to review briefly what is known about $d_1$.
For the pion chiral symmetry dictates $d_{1,\pi}^Q=-M_{2,\pi}^Q$ 
\cite{Polyakov:1999gs,Teryaev:2001qm}.
For the nucleon the large-$N_c$ limit predicts the flavour-dependence
$|d_1^u+d_1^d| \gg |d_1^u-d_1^d|$ \cite{Goeke:2001tz} which is supported by
lattice calculations \cite{Hagler:2003jd,Gockeler:2003jf,Negele:2004iu} 
and results from the chiral quark soliton model 
\cite{Wakamatsu:2006dy,Goeke:2007fp,Goeke:2007fq,Petrov:1998kf,Kivel:2000fg,Schweitzer:2002nm}. 
Both lattice QCD and chiral quark soliton model yield a negative 
$d_1^Q=d_1^u+d_1^d$. Also for nuclei $d_1^Q$ is found to be negative
\cite{Polyakov:2002yz,Guzey:2005ba}.
In Ref.~\cite{Goeke:2007fp} it was conjectured on the basis of plausible physical
arguments that the negative sign of $d_1$ is dictated by stability criteria.

The value of $d_1$ is of interest for phenomenology, since it contributes to 
the beam charge asymmetry in deeply virtual Compton scattering \cite{Kivel:2000fg}.
HERMES data \cite{Ellinghaus:2002bq} indicate a $d_1^Q$ of negative sign though 
this observation depends to some extent on the employed model for the behaviour 
of generalized parton distributions in the small-$x$ region \cite{Belitsky:2001ns}.

\section{The Skyrme model}
\label{Sec-3:model}

In this Section we introduce briefly the Skyrme model.
The Lagrangian of the Skyrme model is given by
\be\label{Eq:Lagrangian}
    {\mathcal L}=
        \frac{F_\pi^2}{16}\;{\rm tr_F}(\partial_\mu U\partial^\mu U^\dag)
    +  \frac{1}{32e^2} \;{\rm tr_F}
    [U^\dag(\partial_\mu U),U^\dag(\partial_\nu U)]\,
    [U^\dag(\partial^\mu U),U^\dag(\partial^\nu U)]
    +  \frac{m_\pi^2F_\pi^2}{8}\;{\rm tr_F}(U-2)
\ee
where $U$ is the SU(2) pion field, $m_\pi$ is the pion mass,
$e$ is a dimensionless parameter, ${\rm tr_F}$ denotes the trace
over the flavour-SU(2) matrices, and $F_\pi$ is the pion decay constant,
with $F_\pi=186\,{\rm MeV}$ in nature. The parameters scale as
\be\label{Eq:parameters-large-Nc}
    F_\pi = {\mathcal O}(N_c^{1/2}) \, , \;\;\;
        e = {\mathcal O}(N_c^{-1/2})    \, , \;\;\;
    m_\pi = {\mathcal O}(N_c^0)     \, ,
\ee
in the large $N_c$ limit, such that ${\mathcal L}={\mathcal O}(N_c)$
which guarantees the correct large-$N_c$ scaling of the theory.

In the large-$N_c$ limit the chiral field $U$ is static.
Using the Skyrme (or ``hedgehog'') Ansatz, $U=\exp[i\btau{\bf e_r} F(r)]$
with $r=|{\bf x}|$ and ${\bf e_r}={\bf x}/r$, one obtains the soliton mass
$M_{\rm sol}=-\int\di^3{\bf x}\;{\mathcal L}$ as functional of the radial function $F(r)$
\be\label{Eq:Esol}
    M_{\rm sol}[F]
  = 4\pi\int\limits_0^\infty\di r\; r^2\biggl[
    \frac{F_\pi^2}{8}\biggl(\frac{2\sin^2F(r)}{r^2}+F^\prime(r)^2\biggr)
  + \frac{\sin^2F(r)}{2\,e^2\,r^2}\biggl(\frac{\sin^2F(r)}{r^2}+2F^\prime(r)^2\biggr)
  + \frac{m_\pi^2F_\pi^2}{4}\,\biggl(1-\cos F(r)\biggr) \biggr]\;. \ee
The soliton profile $F(r)$ which minimizes (\ref{Eq:Esol}) is determined by
the following differential equation
\be\label{Eq:diff-eq}
    \biggl(\frac{r^2}{4}+\frac{2\sin^2F(r)}{e^2\,F_\pi^2}\biggr)F^{\prime\prime}(r)
    +\frac{rF^\prime(r)}{2}
    +\frac{F^\prime(r)^2\,\sin 2F(r)}{e^2 F_\pi^2}
    -\frac{\sin 2F(r)}{4}
    -\frac{\sin^2 F(r)\,\sin 2F(r)}{e^2 F_\pi^2 r^2}
    -\frac{m_\pi^2r^2\sin F(r)}{4} = 0\,  \ee
with the boundary conditions $F(0)=\pi$ and $F(r)\to 0$ with $r\to\infty$.
These conditions ensure that the soliton field has unity winding number
which is identified with the baryon number in the Skyrme model.

In order to ascribe spin and isospin quantum numbers to the soliton one
considers $U({\bf x}) \to A(t)U({\bf x})A^{-1}(t)$ in (\ref{Eq:Lagrangian})
with an arbitrary SU(2) matrix $A = a_0+i\,{\bf a}${\boldmath $\tau$},
introduces conjugate momenta $\pi_b=\partial L/\partial\dot{a}_b$, and
quantizes the collective coordinates according to
$\pi_b \to -i \partial /\partial a_b$ considering the constraint
$a_0^2+{\bf a}^2=1$. This yields the Hamiltonian
\be\label{Eq:H-rot}
    H = M_{\rm sol} + \frac{{\bf J}^2}{2\Theta}
          = M_{\rm sol} + \frac{{\bf I}^2}{2\Theta}
\ee
where ${\bf J}^2$ and ${\bf I}^2$ are the squared spin and isospin operators,
respectively. These operators act on the nucleon or $\Delta$ wave functions
which can be expressed in terms of the collective coordinates $a_i$,
for example, the wave function for a proton with spin up is given by
$|p^\uparrow\ra=(a_1+ia_2)/\pi$, etc.\  \cite{Adkins:1983ya}.
Here $\Theta$ denotes the soliton moment of inertia
\be\label{Eq:mom-ienertia}
    \Theta = \frac{2\pi}{3}\int\limits_0^\infty\di r\;
    r^2 \sin^2F(r)\biggl[F_\pi^2+\frac{4 F^\prime(r)^2}{e^2}
    +\frac{4 \sin^2F(r)}{e^2r^2}\biggr]\;.
\ee

It is customary to treat $F_\pi$ and $e$ as free parameters.
One way to fix them consists in adjusting $F_\pi$ and $e$ such
that the masses of nucleon and $\Delta$-resonance are reproduced
\cite{Adkins:1983ya,Adkins:1983hy}.
Hereby one first determines the function $F(r)$ which minimizes
$M_{\rm sol}$ and then projects on the proper quantum numbers.
I.e.\  one considers the nucleon and $\Delta$-resonance
(and hyperons in the SU(3) version of the model)
as different rotational excitations of the same object.

In order to guarantee a {\sl consistent} description of the EMT form factors we 
proceed as follows: We minimize $M_{\rm sol}$ and quantize collective coordinates
as described above. In general the model expression for any observable $A$ is 
\be\label{Eq:general-observable}
      A = A_{\rm LO} + A_{\rm rot}\;,
\ee
where $A_{\rm LO}$ appears in leading order of the large-$N_c$ expansion,
while $A_{\rm rot}$ arises from soliton rotations and is suppressed by $1/N_c$ 
with respect to $A_{\rm LO}$. For symmetry reasons $A_{\rm LO}$ could vanish.
If $A_{\rm LO}\neq 0$ we shall neglect the contribution of $A_{\rm rot}$
(unless it happens to vanish anyway). In other words, we consider corrections 
due to soliton rotation if and only if the leading order gives a vanishing result.
This step is consistent from a rigorous large-$N_c$ limit point of view,
but it implies that we strictly speaking have to limit ourselves to the comparison 
of isoscalar and separately isovector flavour combinations of nucleon properties, 
see Table.~\ref{Table-I}, which are in general of different order in the large-$N_c$ 
expansion.
The results obtained below for the total EMT form factors refer, however, both 
to the proton and neutron.
The reasons why other approaches are not favourable are discussed 
in App.~\ref{App:rot-corr}.

The masses of the nucleon and $\Delta$-resonance as they follow from
(\ref{Eq:H-rot}) are of the type (\ref{Eq:general-observable}).
(In this case $A_{\rm rot}$ is suppressed with respect to
$A_{\rm LO}$ by two orders in $1/N_c$.)
In our approach $M_N$ and $M_\Delta$ are degenerated and
\be\label{Eq:Mn}
    M_N=M_\Delta=M_{\rm sol}\equiv \min\limits_F M_{\rm sol}[F]\;.
\ee
Experimentally the nucleon and $\Delta$ masses differ by about $30\%$
which indicates the typical accuracy we may expect (in the best case)
for our results. The Skyrme model is, in fact, observed to agree with
phenomenology to within a similar accuracy \cite{Holzwarth:1985rb}. 
From this point of view our approach is ``within the accuracy of the model''.

In order to fix the parameters $F_\pi$ and $e$ we choose to
reproduce exactly the following observables\footnote{
        \label{Footnote:parameter}
        Also other choices are possible. In this way, however, we preserve the 
	``tradition'' \cite{Adkins:1983ya,Adkins:1983hy} of using baryon masses 
	for parameter fixing, and obtain a comparably satisfactory description 
	of the nucleon, see Table~\ref{Table-I}. 
	We stress that in our approach Eq.~(\ref{Eq:para-fixing}) 
	does not imply that $M_N$ and $M_\Delta$ are correctly reproduced.
	Instead, due to (\ref{Eq:Mn}), one has 
	$M_N=M_\Delta=1085\,{\rm MeV}$ 	
	which agrees to within $15\,\%$ with the experimental values. 
	Notice that the nucleon mass appears overestimated
	also in other soliton models.
	The origin of this problem is understood, and 
	one way to solve it was discussed in \cite{Meier:1996ng} where
	it was demonstrated that quantum corrections due to meson fluctuations 
	significantly reduce the soliton energy. 
	A different way of understanding this problem, namely on the basis of 
	notions and methods known from nuclear physics, was indicated in 
	\cite{Pobylitsa:1992bk}.}
\be\label{Eq:para-fixing}
    M_\Delta+M_N\stackrel{!}{\equiv}2M_{\rm sol}=2171\,{\rm MeV}\;,\;\;\;
    M_\Delta-M_N\stackrel{!}{\equiv}\frac{3}{2\Theta}=293\,{\rm MeV}\;.
\ee
This requires for $m_\pi=138\,{\rm MeV}$ the following values
\be\label{Eq:para-fixed}
    F_\pi = 131.3\,{\rm MeV} \;,\;\;\; e = 4.628\,.
\ee
The experimental value of $F_\pi=186\,{\rm MeV}$ is underestimated by
$30\,\%$ which is typical in this model. In order to gain intuition
on the performance of the model with parameters fixed according to
(\ref{Eq:para-fixing},~\ref{Eq:para-fixed})
we consider several nucleon observables:
isoscalar ($I=0\equiv$ proton $+$ neutron) and
isovector ($I=1\equiv$ proton $-$ neutron) electric and magnetic
charge square radii $\la r_{el}^2\ra_I$ and $\la r_m^2\ra_I$,
magnetic moments $\mu_I$, the axial coupling constant $g_A$,
and pion-nucleon sigma-term $\sigma_{\pi N}$. 
The latter is evaluated analogously to Ref.~\cite{Adkins:1983hy}
yielding $\sigma_{\pi N}=84\,{\rm MeV}$, which agrees well with 
results from recent analyses \cite{recent-sigma-extractions},
see Table~\ref{Table-I}, which indicate a larger value for $\sigma_{\pi N}$
than earlier extractions \cite{earlier-sigma-extractions}, and also with
results from other models, see \cite{Schweitzer:2003sb}.
All results are summarized in Table~\ref{Table-I}, and agree with
phenomenology to within an accuracy of $30\%$ with the exception of $g_A$
which, however, comes close to its estimated large-$N_c$ value
\cite{Jackson:1983bi}.

%
\begingroup
\squeezetable
\begin{table}[b!]
    \caption{\footnotesize\sl
    \label{Table-I}
     The nucleon observables electric and magnetic charge square radii
     $\la r_{el}^2\ra_I$ and $\la r_m^2\ra_I$, magnetic moments $\mu_I$,
     and the axial coupling constant $g_A$
     obtained in the Skyrme model using baryon masses according to
     Eqs.~(\ref{Eq:para-fixing},~\ref{Eq:para-fixed}),
     see text and Footnote~\ref{Footnote:parameter}.}
\vspace{0.2cm}
    \begin{ruledtabular}
    \begin{tabular}{llll}
    \\
Quantity &
\multicolumn{2}{l}{Skyrme model, results obtained here for} &
Experiment  \cr
 &    $m_\pi = 0$ & $138\,{\rm MeV}$ \cr
     \\
     \hline
     \\
$M_\Delta+M_N$ in MeV            & 2068     & 2171 (fixed) & 2171   \cr
$M_\Delta-M_N$ in MeV        & 183      & 293  (fixed) & 293    \cr
$F_\pi$ in MeV               & 131      & 131         & 186     \cr
$\la r_{el}^2\ra_{I=0}^{1/2}$ in fm & 0.69     & 0.60        & 0.72    \cr
$\la r_{el}^2\ra_{I=1}^{1/2}$ in fm & $\infty$ & 0.96        & 0.88    \cr
$\la r_m^2\ra_{I=0}^{1/2}$ in fm & 1.07     & 0.85        & 0.81    \cr
$\la r_m^2\ra_{I=1}^{1/2}$ in fm & $\infty$ & 0.96        & 0.80    \cr
$\mu_{I=0}$ in nuclear magnetons & 0.51     & 0.66        & 0.88    \cr
$\mu_{I=1}$ in nuclear magnetons & 5.66     & 3.70        & 4.70    \cr
$\mu_{I=0}/\mu_{I=1}$        & 0.09     & 0.18        & 0.19    \cr
$g_A$                        & 0.84     & 0.73        & 1.26    \cr
$\sigma_{\pi N}$ in MeV      & 0        & 84          & 60-80   \cr
\cr
\end{tabular}
\end{ruledtabular}
\end{table}
\endgroup
%

It is of interest to consider nucleon observables in the chiral limit.
For that it is necessary to specify what happens to the Skyrme model
parameters $F_\pi$ and $e$ as $m_\pi$ varies. In this work we shall keep
these parameters fixed at their values in the physical situation in
Eq.~(\ref{Eq:para-fixed}).
Hereby we neglect the fact that in general $F_\pi$ depends on $m_\pi$
\cite{Gasser:1983yg,Colangelo:2001df}.
In this way we find that the nucleon (and $\Delta$) mass are reduced in
the chiral limit by
\be\label{Eq:MnPhys-MnChir}
      M_N|_{\mbox{\footnotesize\sl physical point}}\;-\;
      M_N|_{\mbox{\footnotesize\sl chiral limit}}\;
      = 51\,{\rm MeV}\,,
\ee
which is within the range of the values considered in chiral perturbation
theory. The isovector electric and magnetic charge square radii diverge
in the chiral limit. Further results are shown in Table~\ref{Table-I}.

We remark that the description of the nucleon in the Skyrme model with the
parameter fixing (\ref{Eq:para-fixing},~\ref{Eq:para-fixed}) is --- both
for finite $m_\pi$ and in the chiral limit, see Table~\ref{Table-I} ---
comparable to that in Refs.~\cite{Adkins:1983ya,Adkins:1983hy}.

\section{The EMT form factors in the Skyrme model}
\label{Sec-4:EMT-ffs-in-model}

In the Skyrme model the canonical EMT, which --- introducing the notation
$U\equiv \phi_0+i\tau_a\phi_a$ --- is given by
\be\label{Eq:EMT-canon}
    T^{\mu\nu} =
    \frac{\partial\,\mathcal L\;}{\partial(\partial_\mu\phi_a)}\,\partial^\nu\phi_a
    -g^{\mu\nu}\,{\mathcal L}\;,
\ee
is already symmetric.  
(Notice that in (\ref{Eq:EMT-canon}) one has to sum over $a=1,\,2,\,3$; and
$\phi_0$ must not be considered as an independent field, since it is constrained
by $\phi_0^2+\phi_a^2=1$.)
For the respective components of the static EMT
in Eq.~(\ref{Def:static-EMT}) one obtains 
(c.f.\  App.~\ref{App:rot-corr})
\ba     T^{00}(r)
    &=&
        \frac{F_\pi^2}{8}\biggl(\frac{2\sin^2F(r)}{r^2}+F^\prime(r)^2\biggr)
        +\frac{\sin^2F(r)}{2\,e^2\,r^2}\biggl(\frac{\sin^2F(r)}{r^2}+2F^\prime(r)^2
    \biggr)+\frac{m_\pi^2F_\pi^2}{4}\,\biggl(1-\cos F(r)\biggr)
    \label{Eq:T00-Skyrme}\\
        T^{0k}({\bf r},{\bf s})
    &=&
     \frac{\epsilon^{klm}r^ls^m}{({\bf s}\times{\bf r})^2}\;\rho_J(r)
    \;\;\;\mbox{with}\;\;\;
    \rho_J(r)=\frac{\sin^2F(r)}{12\,\Theta}
        \biggl[F_\pi^2+\frac{4 F^\prime(r)^2}{e^2}+\frac{4 \sin^2F(r)}{e^2r^2}\biggr]
    \label{Eq:T0k-Skyrme}
\ea
while $T^{ij}(r)$ is given by Eq.~(\ref{Eq:T_ij-pressure-and-shear}) with
\ba
    \label{Eq:pressure}
    p(r) &=& -\frac{F_\pi^2}{24}\biggl(\frac{2\sin^2F(r)}{r^2}+F^\prime(r)^2\biggr)
    +\frac{\sin^2F(r)}{6e^2\,r^2}\biggl(\frac{\sin^2F(r)}{r^2}+2F^\prime(r)^2
    \biggr)-\frac{m_\pi^2F_\pi^2}{4}\,\biggl(1-\cos F(r)\biggr)\\
    \label{Eq:shear}
    s(r) &=& \biggl(\frac{F_\pi^2}{4}+\frac{\sin^2F(r)}{e^2\,r^2}\biggr)
    \biggl(F^\prime(r)^2-\frac{\sin^2F(r)}{r^2}\biggr)\;.
    \ea
In the large-$N_c$ limit $M_N\sim {\mathcal O}(N_c)$ while the components of
the 4-momentum transfer behave as $\Delta^0\sim {\mathcal O}(N_c^{-1})$ and
$\Delta^i\sim {\mathcal O}(N_c^0)$. Therefore $|t|\ll M_N^2$.
Considering this and the large-$N_c$ relations (\ref{Eq:FF-large-Nc})
one obtains from (\ref{Eq:ff-J},~\ref{Eq:ff-d1},~\ref{Eq:ff-M2})
\ba
        M_2(t)-\frac{t}{5M_N^2}d_1(t)
        &=& \frac{1}{M_N}\int\di^3{\bf r}\;T_{00}(r)\;j_0(r\sqrt{-t})
        \label{Eq:M2-d1-model-comp}\\
        d_1(t)
        &=& \frac{15 M_N}{2}\int\di^3{\bf r}\;p(r) \;\frac{j_0(r\sqrt{-t})}{t}
        \label{Eq:d1-model-comp}\\
        J(t)
        &=& 3
    \int\di^3{\bf r}\;\rho_J(r)\;\frac{j_1(r\sqrt{-t})}{r\sqrt{-t}}\;,
    \label{Eq:J-model-comp}
\ea
with $T_{00}(r)$, $p(r)$ and $\rho_J(r)$ as defined in
Eqs.~(\ref{Eq:T00-Skyrme},~\ref{Eq:T0k-Skyrme},~\ref{Eq:pressure}), and the
Bessel functions $j_0(z) = \frac{\sin z}{z}$ and $j_1(z)=-j_0^\prime(z)$.

From the large-$N_c$ behaviour (\ref{Eq:parameters-large-Nc}) of the
parameters $F_\pi$ and $e$ we see that $M_N={\mathcal O}(N_c)$
and $\Theta={\mathcal O}(N_c)$ such that
$M_2(t)={\mathcal O}(N_c^0)$, $J(t)={\mathcal O}(N_c^0)$ and $d_1(t)={\mathcal O}(N_c^2)$
holds in agreement with the general large-$N_c$ result (\ref{Eq:FF-large-Nc}).

Next we verify the internal consistency of the derived model
expressions and demonstrate that they satisfy the general relations
(\ref{Eq:M2-J-d1},~\ref{Eq:relation-p(r)-s(r)},~\ref{Eq:stability}).
For that we first take the limit $t\to0$ in Eq.~(\ref{Eq:M2-d1-model-comp})
and use the fact that $d_1(t)$ is well-defined at $t=0$, see below.
By considering Eqs.~(\ref{Eq:Esol},~\ref{Eq:Mn}) we see that
the first constraint in Eq.~(\ref{Eq:M2-J-d1}) is satisfied
\be\label{Eq:M2-d1-model-comp-at-t=0}
        M_2(0) = \frac{1}{M_N}\int\di^3{\bf r}\;T_{00}(r) = 1\;.
\ee
Similarly, taking $t\to 0$ in Eq.~(\ref{Eq:J-model-comp}) and comparing
Eqs.~(\ref{Eq:mom-ienertia},~\ref{Eq:T0k-Skyrme}) we see that the second
constraint in Eq.~(\ref{Eq:M2-J-d1}) is satisfied
\be\label{Eq:J-model-comp-at-t=0}
        J(0) = \int\di^3{\bf r}\;\rho_J(r)=\frac12\;.
\ee
To see that also the third constraint in
Eq.~(\ref{Eq:M2-J-d1}) is satisfied it is convenient to prove
first Eqs.~(\ref{Eq:relation-p(r)-s(r)},~\ref{Eq:stability}).
To prove the stability condition (\ref{Eq:stability}) we rewrite
$r^2p(r)$ as (for an alternative proof see App.~\ref{App-stabilty})
\be\label{Eq:proof-stability}
    r^2p(r) = \frac{\partial\;}{\partial r}\biggl[
     r^3p(r)
    +\frac{F_\pi^2}{12}\,r^3F^\prime(r)^2
    -\frac{\sin^4F(r)}{3e^2r}
    +\frac{F_\pi^2m_\pi^2}{6}\,r^3(1-\cos F(r))\biggr]
    -\frac{F_\pi^2}{3}\,rF^\prime(r)
    \times\biggl(\mbox{equations of motion}\biggr)\,,\;\;\;
    \ee
where ``$(\mbox{equations of motion})$'' denotes the left-hand-side of
Eq.~(\ref{Eq:diff-eq}) and vanishes.
Thus, $r^2p(r)$ is a total derivative of a function which is zero at the
boundaries of the integral in Eq.~(\ref{Eq:stability}),
i.e.\  $\int_0^\infty\di r\;r^2p(r) = 0$  is satisfied.
Next, by inserting the expressions (\ref{Eq:pressure}) and (\ref{Eq:shear})
into the differential equation (\ref{Eq:relation-p(r)-s(r)}) we see
--- in the notation of Eq.~(\ref{Eq:proof-stability}) --- that
\be\label{Eq:proof-relation-p(r)-s(r)}
    \frac23\;\frac{\partial s(r)}{\partial r\;}+
        \frac{2s(r)}{r} + \frac{\partial p(r)}{\partial r\;} =
    -\frac{F_\pi^2}{r^2}\,F^\prime(r)
    \times\biggl(\mbox{equations of motion}\biggr) = 0\,.
    \ee
Now we are in the position to check the third constraint in
Eq.~(\ref{Eq:M2-J-d1}) for $d_1(t)$ at zero momentum transfer.
Expanding the expression in (\ref{Eq:d1-model-comp}) for small $t$, one obtains
$\frac{j_0(r\sqrt{-t})}{t}=\frac{1}{t}+\frac{r^2}{3!}+t\frac{r^4}{5!}+{\mathcal O}(t^2)$.
The coefficient of the $\frac{1}{t}$-term in this series is proportional to
$\int_0^\infty\di r\;r^2p(r)$, i.e.\ identically zero. The next term in
this series yields the expression for $d_1$ in terms of $p(r)$,
cf.\  Eq.~(\ref{Eq:d1-from-s(r)-and-p(r)}).
The expression for $d_1$ in terms of $s(r)$ in (\ref{Eq:d1-from-s(r)-and-p(r)})
automatically follows from integrations by parts and the relation
(\ref{Eq:relation-p(r)-s(r)}) which we have shown to be valid in the model,
see Eq.~(\ref{Eq:proof-relation-p(r)-s(r)}).

\section{\boldmath
Properties of the densities and results}
\label{Sec-5:properties-of-the-densities-and-results}

It is instructive to consider first the Fourier transforms the nucleon
EMT form factors (\ref{Eq:M2-d1-model-comp}-\ref{Eq:J-model-comp}),
namely the ``densities''  $T_{00}(r)$, $\rho_J(r)$, $p(r)$ and $s(r)$
in Eqs.~(\ref{Eq:T00-Skyrme}-\ref{Eq:shear}). Some of their
properties can be derived analytically in the Skyrme model.
For example, at small $r$ the densities behave as
\ba
    T_{00}(r)=T_{00}(0)-A\;r^2+{\mathcal O}(r^4)&&\phantom{\frac11}
        \label{Eq:T00-Skyrme-small-r}\\
        p(r)     =\;\;p(0)-\frac13\;B\;r^2+{\mathcal O}(r^4)&&
        \label{Eq:pressure-small-r}\\
        s(r)     =\phantom{XXX+}\frac15\;B\;r^2+{\mathcal O}(r^4)&&
    \label{Eq:shear-small-r}\\
        \rho_J(r)=\phantom{XXXX+}C\;r^2+{\mathcal O}(r^4)&&\phantom{\frac11}
    \label{Eq:rhoJ-Skyrme-small-r}
\ea
which one concludes from solving the differential equation (\ref{Eq:diff-eq})
iteratively at small $r$, see App.~\ref{App:small-r}.
All constants $T_{00}(0)$, $p(0)$ and $A$, $B$, $C$ in
(\ref{Eq:T00-Skyrme-small-r}-\ref{Eq:rhoJ-Skyrme-small-r}) are defined
to be positive numbers, see App.~\ref{App:small-r} for explicit expressions.

Also the large-$r$ behaviour of the densities can be determined
analytically from that of the soliton profile function.
Chiral symmetry and the variational equation (\ref{Eq:diff-eq})
dictate the large distance behaviour of the latter to be as follows
\be\label{Eq:prof-large-r}
    F(r) \stackrel{{\rm large}\,\,r}{=}
    \frac{2R_0^2}{r^2}\,(1+m_\pi r)\,\exp(-m_\pi r) \, , \,\,\;\;
\ee
In the chiral limit the constant $R_0^2$, which is introduced here for later 
convenience, is related to the axial coupling constant $g_A$ according to\footnote{
	The relation of $g_A$ to the asymptotics of the profile (\ref{Eq:prof-large-r})
	is valid only in the chiral limit. For example, for the parameters used here
	Eq.~(\ref{Eq:prof-large-r-II}) would imply for $g_A$ the value $0.76$ 
	to be compared with the correct result $0.73$, see 
	Table~\ref{Table-I}, which follows from evaluating directly the
	axial current matrix element. We are grateful to the referee for 
	drawing our attention to this fact.}
\be\label{Eq:prof-large-r-II}
    R_0^2 = \frac{3\,g_A}{4\pi F_\pi^2}       \, .
\ee

Due to (\ref{Eq:prof-large-r}) the densities decay at large $r$ exponentially
for $m_\pi>0$, but exhibit power-like decays in the chiral limit
\ba
        \label{Eq:T00-Skyrme-large-r}
    T_{00}(r) &=&\; 3R_0^2\;\frac{1}{r^6} + \dots\\
    \label{Eq:pressure-large-r}
    p(r)      &=&   -R_0^2\;\frac{1}{r^6} + \dots\\
    \label{Eq:shear-large-r}
    s(r)      &=&\; 3R_0^2\;\frac{1}{r^6} + \dots\\
    \label{Eq:rhoJ-Skyrme-large-r}
    \rho_J(r) &=&\;\,\frac{R_0^2}{3\Theta}\;\frac{1}{r^4} + \dots
\ea
where the dots denote terms which are more strongly suppressed at large $r$.
The results (\ref{Eq:T00-Skyrme-large-r}-\ref{Eq:rhoJ-Skyrme-large-r})
allow to draw several interesting conclusions. Let us define the mean
square radii of the energy and angular momentum density as
\ba
        \label{Eq:def-energy-mean-square-radius}
    \la r_E^2\ra &=&
    \frac{\int\di^3{\bf r}\;r^2 T_{00}(r)}{\int\di^3{\bf r}\;T_{00}(r)}\;,\\
        \label{Eq:def-ang-mom-mean-square-radius}
    \la r_J^2\ra &=&
    \frac{\int\di^3{\bf r}\;r^2 \rho_J(r)}{\int\di^3{\bf r}\;\rho_J(r)}\;.
\ea
We see that in the chiral limit: $\la r_E^2\ra$ is finite, $\la r_J^2\ra$ diverges,
$d_1\propto\int_0^\infty\di r\,r^4 p(r)$ is finite, but the slope of $d_1(t)$ at
zero momentum transfer, $d_1^\prime(0)\propto\int_0^\infty\di r\,r^6p(r)$,
diverges.

The same small- and large-$r$ behaviour of the densities was observed
in the chiral quark soliton model in \cite{Goeke:2007fp,Goeke:2007fq}. 
(For the latter this is expected because
(\ref{Eq:T00-Skyrme-large-r}-\ref{Eq:rhoJ-Skyrme-large-r}) follow from
the chiral behaviour of the soliton profile (\ref{Eq:prof-large-r})
equally respected in both models.) This indicates that the properties
(\ref{Eq:T00-Skyrme-small-r}-\ref{Eq:rhoJ-Skyrme-small-r}) and
(\ref{Eq:T00-Skyrme-large-r}-\ref{Eq:rhoJ-Skyrme-large-r})
are general features in the large-$N_c$ chiral soliton approach
and independent of dynamical details of underlying theories.

\begin{figure*}[t!]
\begin{tabular}{ccc}
\hspace{-1.0cm}\includegraphics[height=4.1cm]{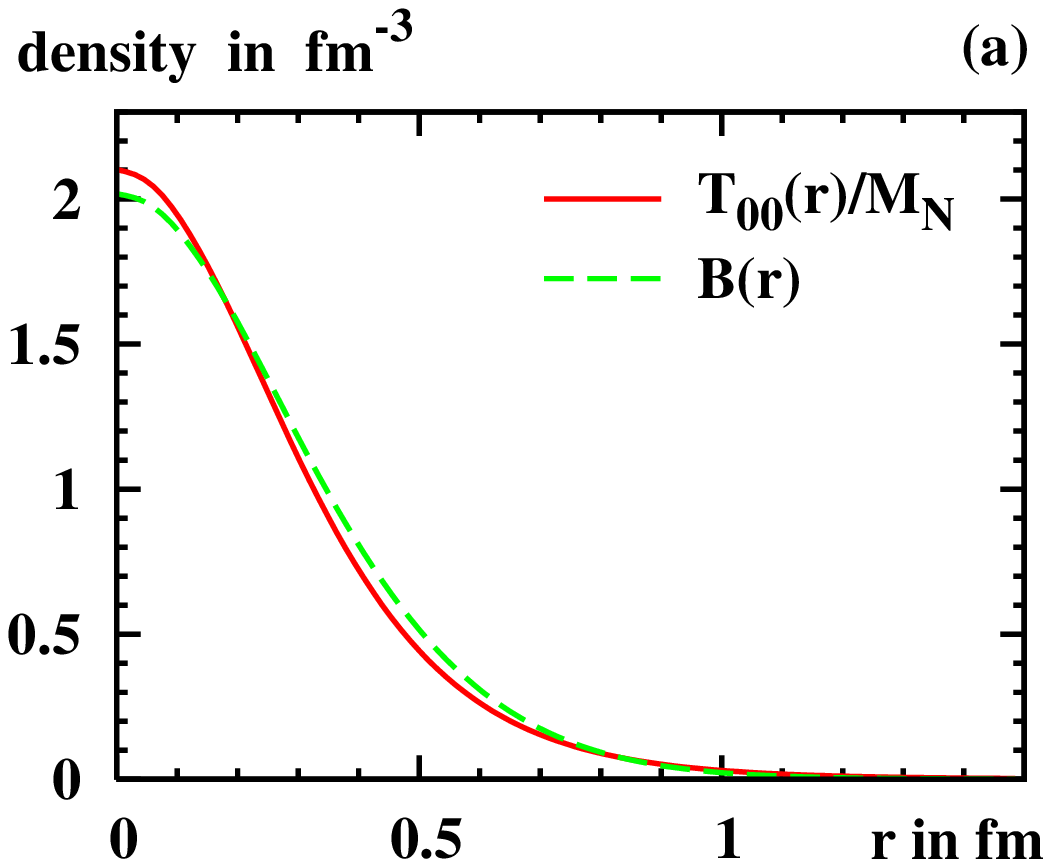}&
\hspace{-0.3cm}\includegraphics[height=4.1cm]{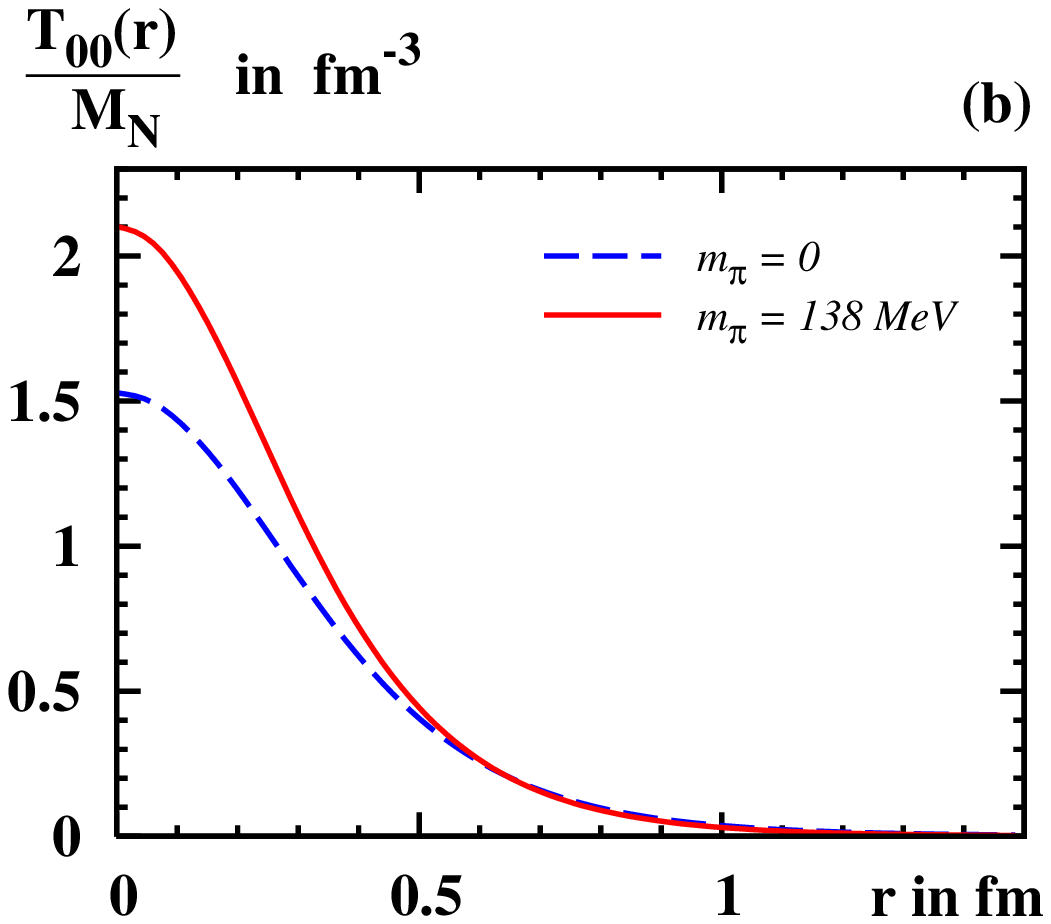}&
\hspace{-0.3cm}\includegraphics[height=4.1cm]{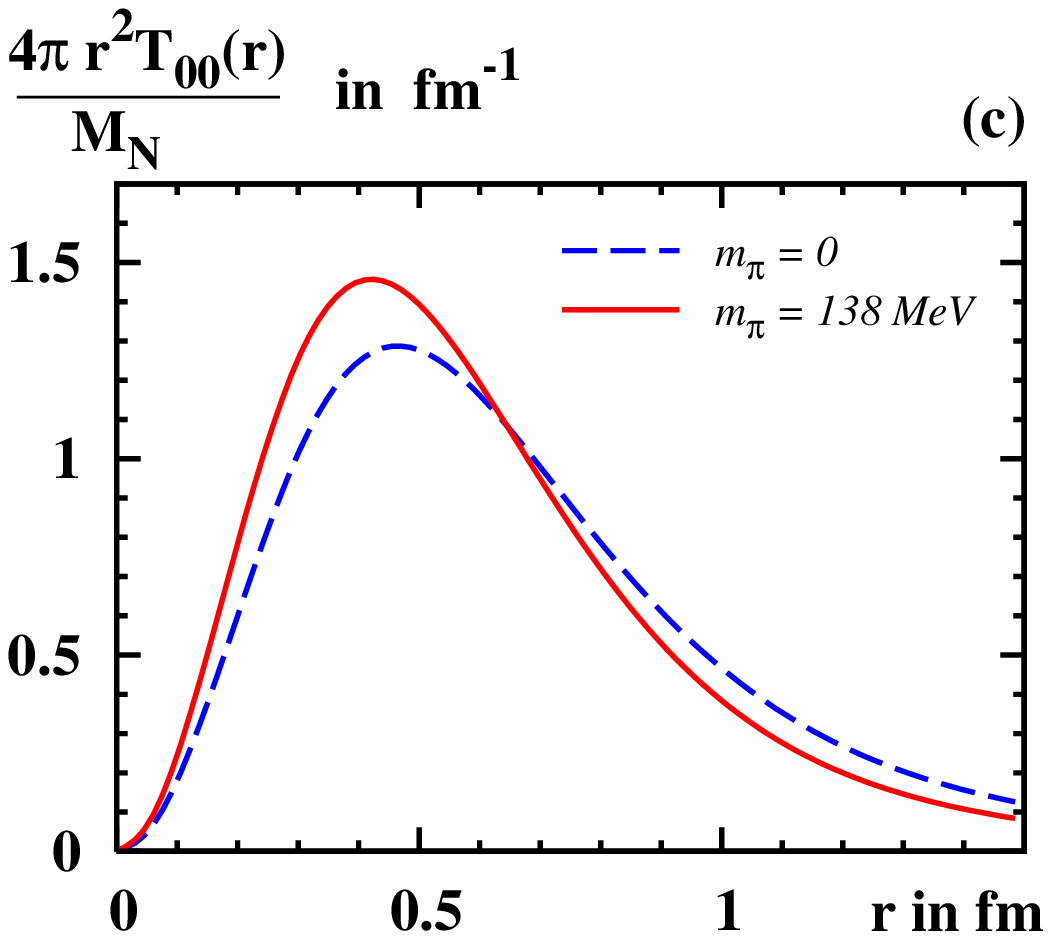}
\end{tabular}
\vspace{-0.4cm}
    \caption{\label{Fig1-energy-density}
    \footnotesize\sl
    (a) The energy density in the nucleon normalized with respect to the
        nucleon mass, $T_{00}(r)/M_N$, as function of $r$ in the Skyrme model
        (solid line), and the baryon density $B(r)$ for sake of comparison
        (dashed line). Both curves yield unity upon integration over the
        full space.
    (b) $T_{00}(r)/M_N$ as function of $r$ in the Skyrme model
        in the physical situation (solid line),
        and in the chiral limit (dashed line).
    (c) The same as (b) but for the normalized energy density
        $4\pi r^2T_{00}(r)/M_N$, i.e.\ here the curves yield unity
        upon integration over $r$.}
\vspace{0.3cm}
\begin{tabular}{cc}
    \includegraphics[height=4.8cm]{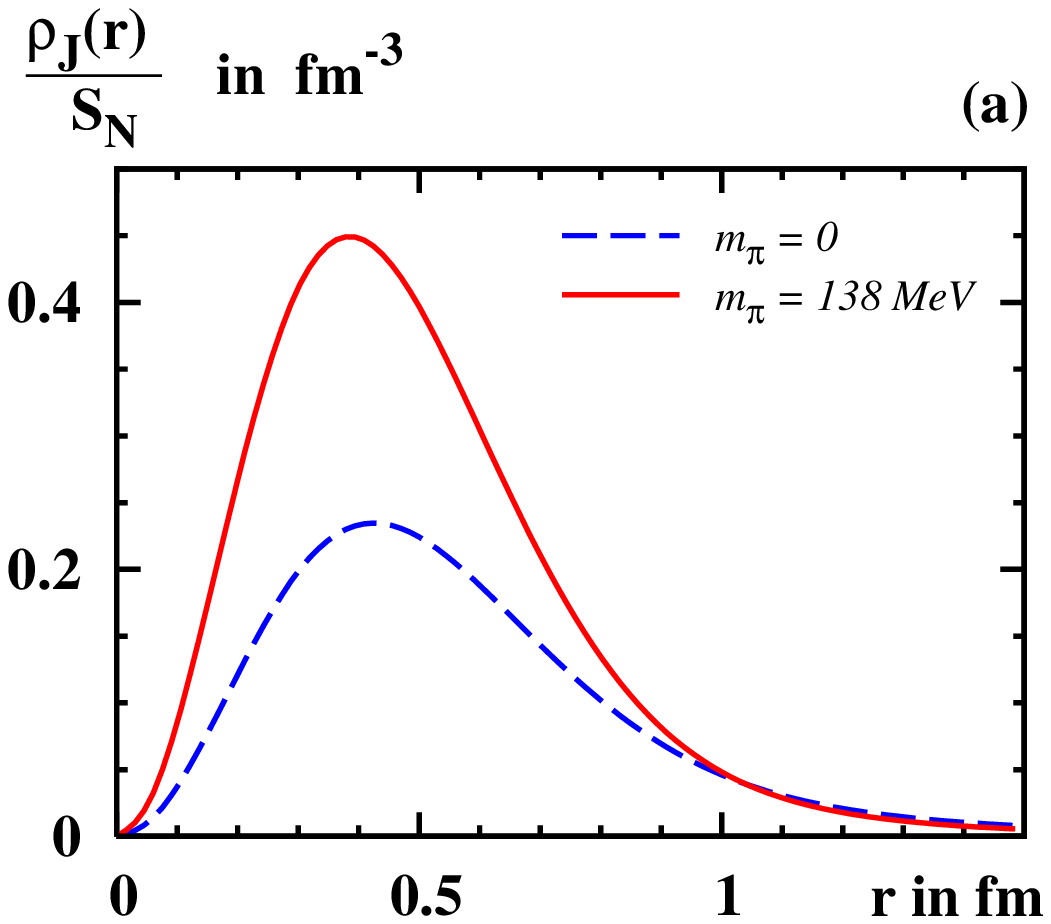}&
    \includegraphics[height=4.8cm]{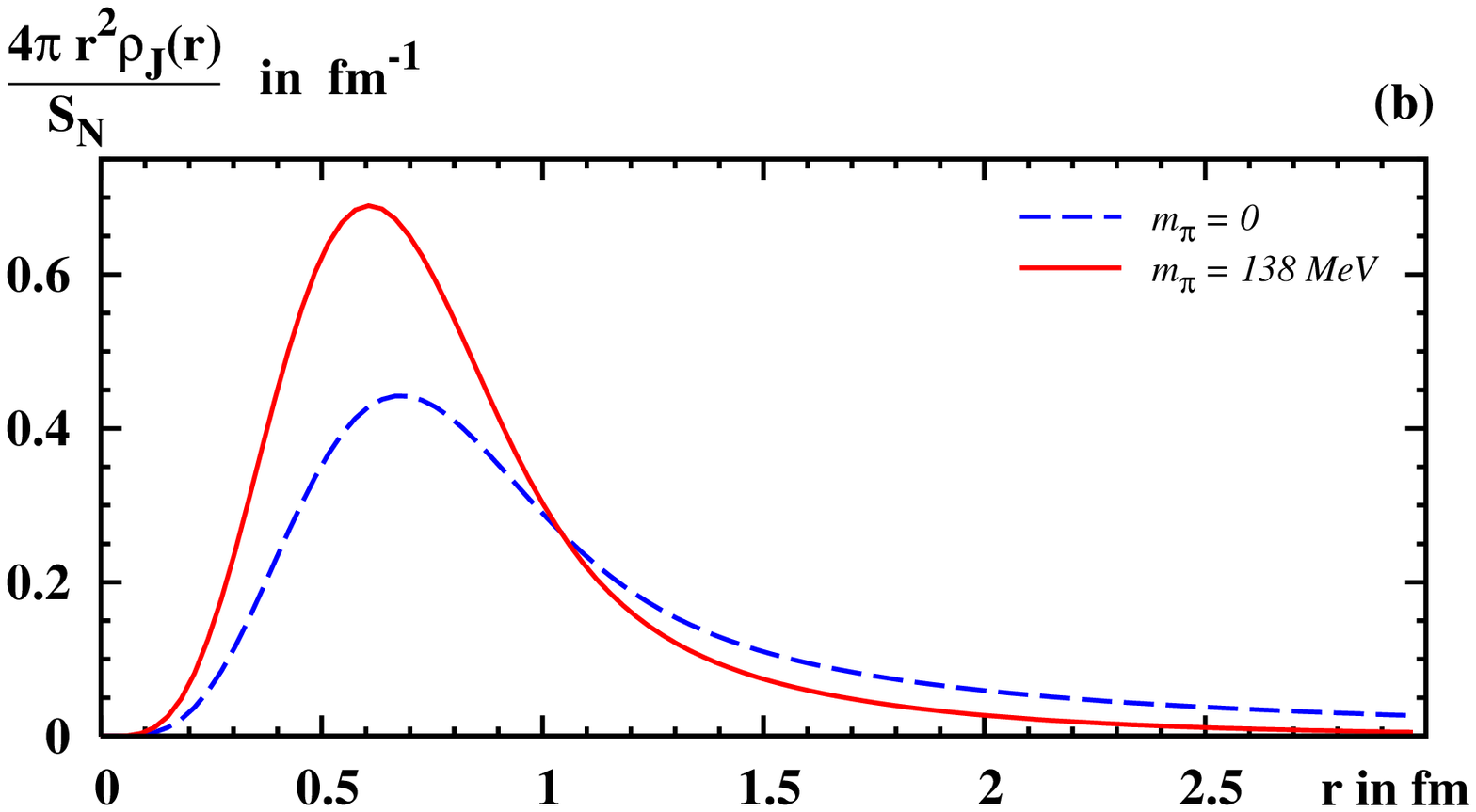}
\end{tabular}
\vspace{-0.4cm}
    \caption{\label{Fig2-spin-density}
    \footnotesize\sl
    (a) The angular momentum density $\rho_J(r)$ normalized with
        respect to the nucleon spin $S_N=\frac12$ as function of $r$
        for the physical pion mass (solid line), and in the chiral limit
        (dashed line). The curves yield unity when integrated over the
        full space.
    (b) The same as (a) but for the normalized  angular momentum density
        $4\pi r^2\rho_J(r)/S_N$, i.e.\ here the curves yield unity
        upon integration over $r$.}
\vspace{0.3cm}
\begin{tabular}{cc}
    \includegraphics[height=4.8cm]{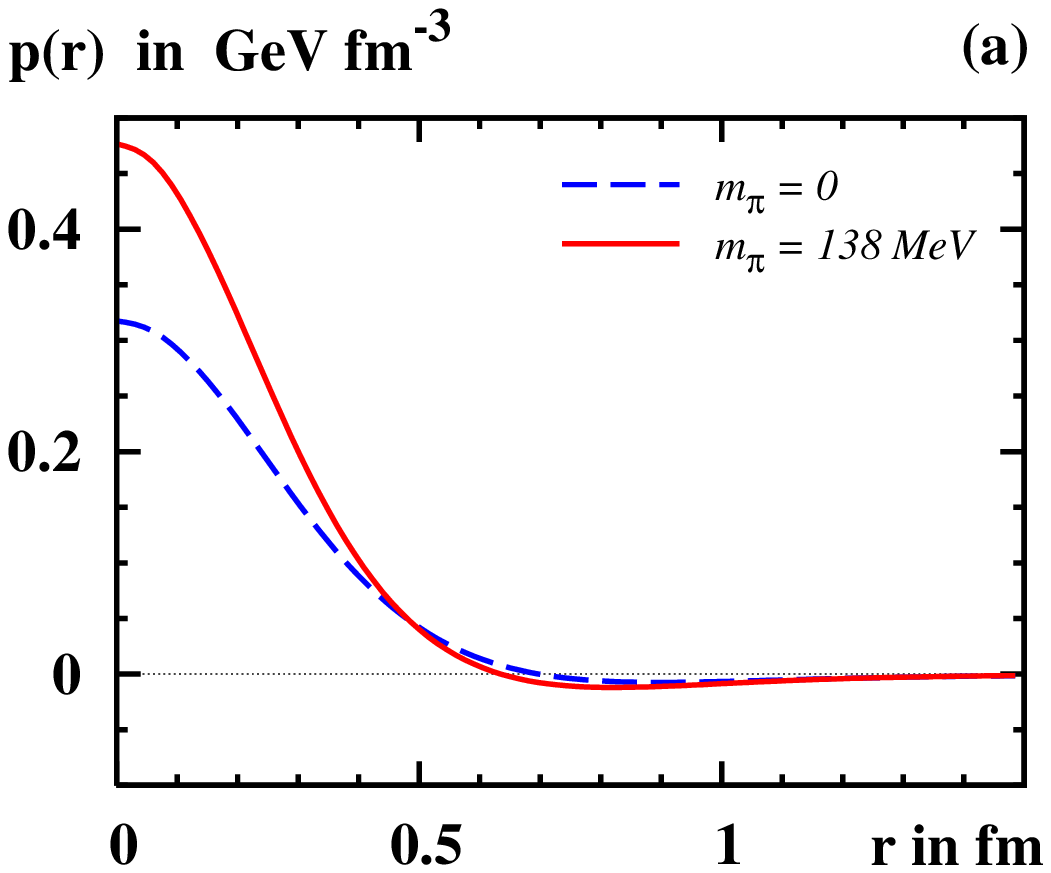}&
    \includegraphics[height=4.8cm]{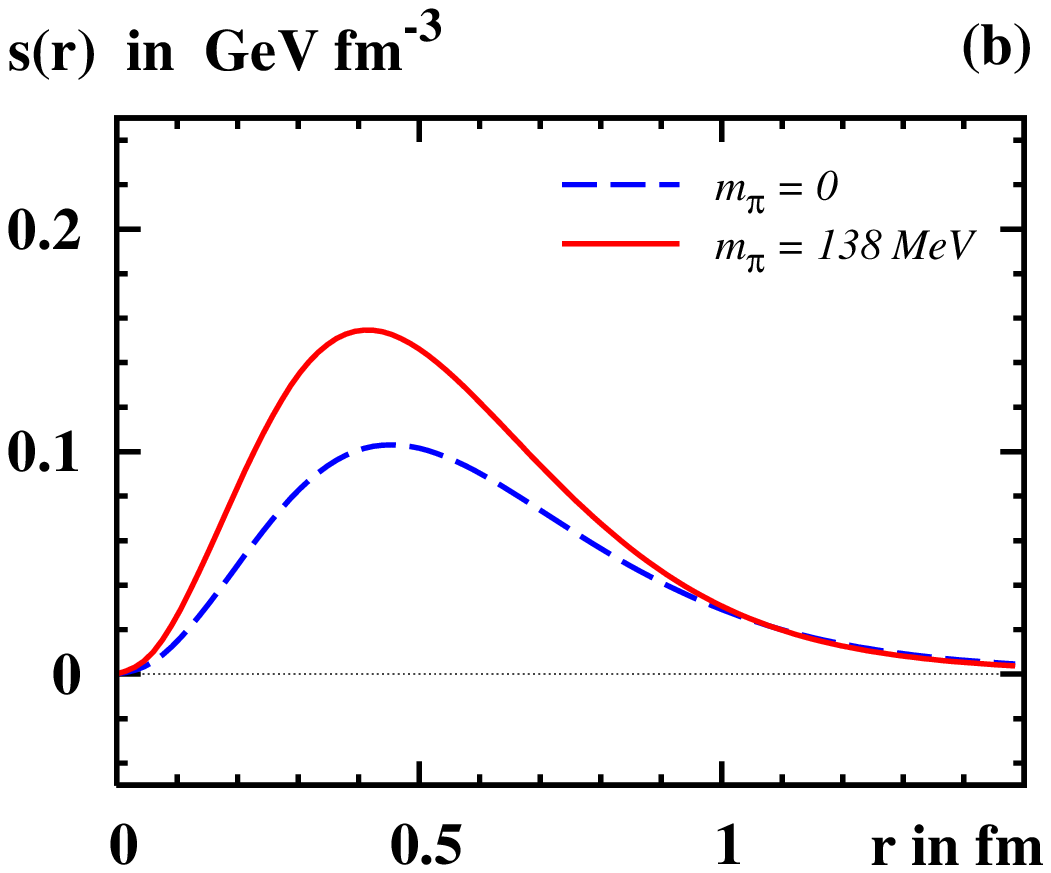}
\end{tabular}
\vspace{-0.4cm}
    \caption{\label{Fig4-pressure+shear}
    \footnotesize\sl
    The distributions of (a) pressure $p(r)$ and (b) the shear forces $s(r)$
    as functions of $r$ in the Skyrme model for the physical pion mass
    (solid line) and in the chiral limit (dashed line).}
\vspace{0.6cm}
    \includegraphics[height=4.8cm]{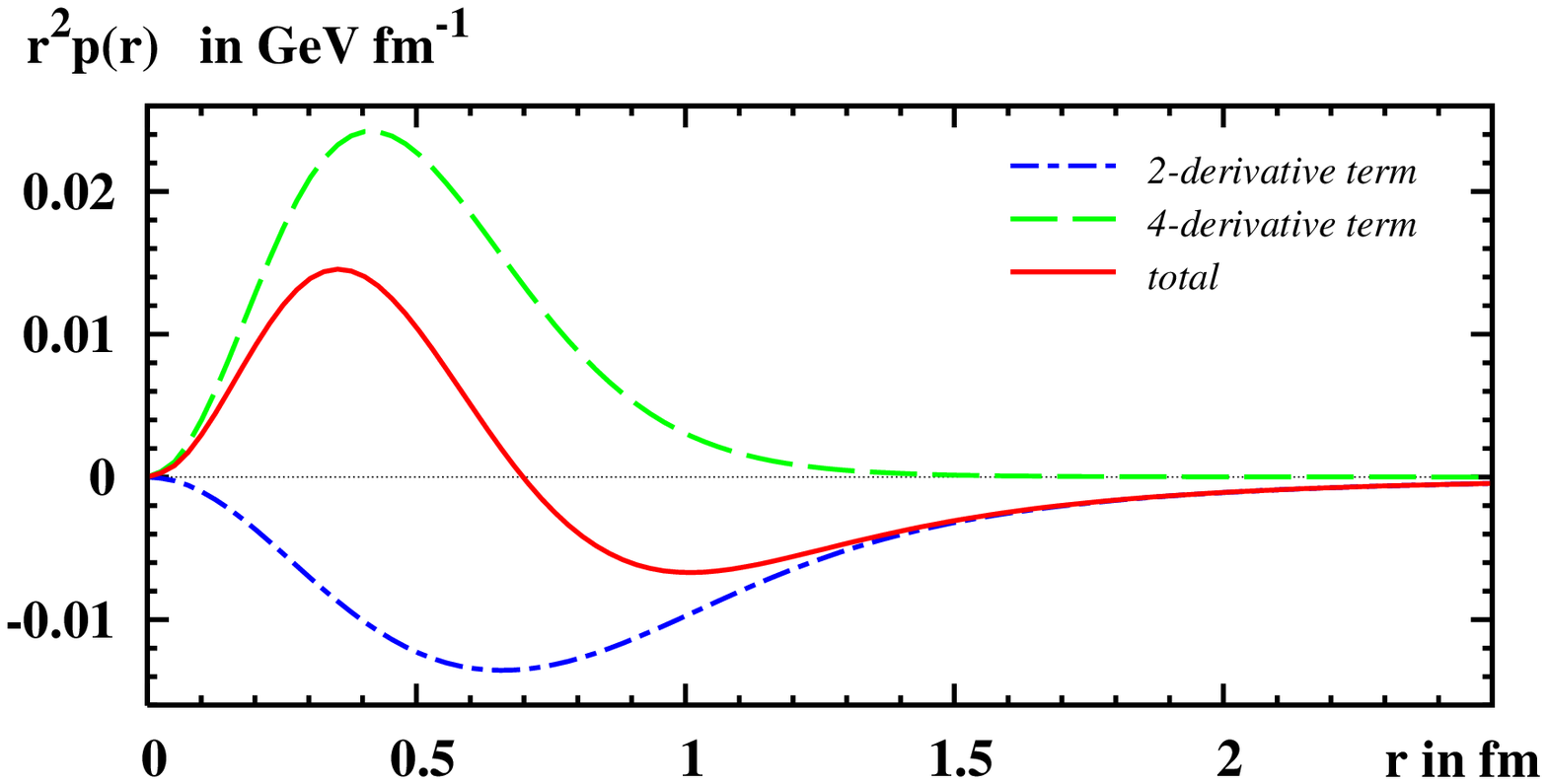}
\vspace{-0.4cm}
    \caption{\label{Fig4-stability}
    \footnotesize\sl
    The decomposition of the pressure in the chiral limit in Skyrme model.
    The graph shows the contributions to $r^2p(r)$ as functions of $r$
    due to the 2-derivative (dashed) and 4-derivative term (dashed-dotted)
    in the Lagrangian (\ref{Eq:Lagrangian}), and the total result
    (solid line).}
\end{figure*}

Next we discuss the numerical results for the densities starting with the
energy density $T^{00}(r)$ of static EMT  (\ref{Def:static-EMT}).
Fig.~\ref{Fig1-energy-density}a shows $T_{00}(r)$ normalized with respect
to the nucleon mass as function of $r$ for the physical pion mass
in comparison to the baryon density $B(r)$ which coincides
with the isoscalar electric charge distribution in the Skyrme model.
Both curves in Fig.~\ref{Fig1-energy-density}a have the same normalization
and yield unity when integrated over the full space.
The similarity of the curves for $r\lesssim 1\,{\rm fm}$ means that
nucleon mass and baryon charge are distributed similarly in this region.
At large $r$, however, $B(r)$ falls off much more rapidly than $T_{00}(r)$,
in the chiral limit $B(r)\propto 1/r^9$ vs.\ Eq.~(\ref{Eq:T00-Skyrme-large-r}).
Therefore  the mean square radii, $\la r_E^2\ra=(0.73\,{\rm fm})^2$
for the physical pion mass (and $(0.94\,{\rm fm})^2$ in  the chiral limit)
in Table~\ref{Table-II}, are much larger than the corresponding values
of $\la r_{el}^2\ra_{I=0}$ in Table~\ref{Table-I}.
The energy density in the center of the nucleon is $2.3\,{\rm GeV\,fm}^{-3}$,
i.e.\  roughly 10 times the equilibrium density of nuclear matter.
Figs.~\ref{Fig1-energy-density}b and c show the behaviour of $T_{00}(r)$
in the chiral limit compared to the physical case with emphasis on
the small- and large-$r$ regions.

Fig.~\ref{Fig2-spin-density} shows the results for the angular momentum
density $\rho_J(r)$ which is related to the $T^{0k}$ components of the
static EMT (\ref{Def:static-EMT}) according to Eq.~(\ref{Eq:T0k-Skyrme}).
In the chiral limit $\rho_J(r)$ is strongly decreased in the small-$r$
region compared to the physical situation, see Fig.~\ref{Fig2-spin-density}a.
This is compensated by the slower decay of $\rho_J(r)$ in the large-$r$
region, see Fig.~\ref{Fig2-spin-density}b, which ensures in either case
the correct normalization $\int\di^3{\bf r}\,\rho_J(r)=S_N=\frac12$,
cf.\ Eq.~(\ref{Eq:M2-J-d1}).
For the mean square radius $\la r_J^2\ra$ of the angular momentum
density (\ref{Eq:def-ang-mom-mean-square-radius})
we obtain $\la r_J^2\ra^{1/2}=0.96\,{\rm fm}$ at the physical point.
In the chiral limit $\la r_J^2\ra$ diverges, see above.

Finally we discuss the distributions of pressure and shear forces
which characterize the $T^{ik}$ components of the static EMT
according to (\ref{Eq:T_ij-pressure-and-shear}).
The pressure, see Fig.~\ref{Fig4-pressure+shear}a,
assumes at $r=0$ its global maximum with\footnote{
  This corresponds to $8\cdot 10^{34}\,{\rm Newton}/{\rm m}^2$.
  One may gain some intuition on this number by considering that
  this is about an order of magnitude higher than the pressure in the
  center of a pulsar according to typical neutron star models.
  By multiplying $p(0)$ with a typical hadronic area of, say, $1\,{\rm fm}^2$
  one obtains a force $\sim 0.5\,{\rm GeV\,fm}^{-1}$ which is about
  half of the QCD string tension.}
$p(0)=0.48\,{\rm GeV\,fm}^{-3}$. It decreases with increasing $r$,
initially according to Eq.~(\ref{Eq:pressure-small-r}), changes
the sign at $r_0=0.64\,{\rm fm}$ till reaching its global minimum
$p_{\rm min}=-0.012{\rm GeV\,fm}^{-3}$ at $r_{\rm min}=0.82\,{\rm fm}$.
For $r>r_{\rm min}$ the pressure increases tending to zero
but remaining always negative according to (\ref{Eq:pressure-large-r}).
The distribution of shear forces, shown in Fig.~\ref{Fig4-pressure+shear}b,
satisfies $s(r)\ge 0$ and becomes zero only for $r\to 0$ and $r\to\infty$
according to Eqs.~(\ref{Eq:shear-small-r},~\ref{Eq:shear-large-r}).
The shear forces exhibit one maximum at $r_{\rm max}=0.41\,{\rm fm}$
with $s_{\rm max}=0.15\,{\rm GeV\,fm}^{-3}$.
The above-mentioned numbers refer to $m_\pi=138\,{\rm MeV}$.
In the chiral limit the distributions are qualitatively similar,
see Fig.~\ref{Fig4-pressure+shear}.

The positive sign of the pressure for $r<r_0$ corresponds to repulsion,
while the negative sign in the region $r > r_0$ means attraction.\footnote{
  Studies in the chiral quark soliton model \cite{Goeke:2007fp} suggest the
  following interpretation. The region $r<r_0$ is dominated by the
  ``quark core'' where Pauli principle is responsible for repulsion.
  The region $r>r_0$ is dominated by the pion cloud responsible for
  binding forces in the effective chiral theory, and thus attraction.
  Due to the lack of explicit quark degrees of freedom no such interpretation
  is possible in the Skyrme model but the results are in qualitative agreement
  with \cite{Goeke:2007fp}.}
In a mechanically stable object the repulsive forces in the inner region
must precisely be balanced by the attractive forces in the outer region ---
in order to satisfy the stability criterion (\ref{Eq:stability}).
In Sec.~\ref{Sec-4:EMT-ffs-in-model} an analytical proof was given that
the condition (\ref{Eq:stability}) is satisfied. It is instructive,
however, to investigate in more detail how the stability is realized.
For that we decompose the pressure in Eq.~(\ref{Eq:pressure}) into the parts
which originate respectively from the 2- and 4-derivative term in the
Lagrangian (\ref{Eq:Lagrangian}) --- considering for simplicity
the chiral limit where the mass term drops out.
(The contribution of the 4-derivative term is labelled
in Eq.~(\ref{Eq:pressure}) by the factor $1/e^2$.)

The contribution of the 2-derivative term is negative for all $r$,
see Fig.~\ref{Fig4-stability}, which means that this term provides
only attraction. This is an alternative way of stating that there is no
soliton solution in the limit $1/e^2\to 0$ in Eq.~(\ref{Eq:Lagrangian}).
The soliton collapses (shrinks to zero) in this limit as can be shown
by means of scaling arguments, see e.g.\   \cite{Holzwarth:1985rb}.

The 4-derivative (Skyrme-) term --- introduced, in fact,
for the sake of stabilizing the soliton --- is always positive,
see Fig.~\ref{Fig4-stability}, i.e.\  responsible for repulsion.
For $r<r_0$ the repulsive forces from the 4-derivative term dominate
over the attractive forces of the 2-derivative term. For $r>r_0$
the situation is reversed. The two terms yield contributions to the
stability condition (\ref{Eq:stability}) which are of opposite sign
\be\label{Eq:stability-in-detail-0}
   \int\limits_0^\infty\di r\;r^2p(r)|_{\mbox{\footnotesize 4-derivative term}}
 =-\int\limits_0^\infty\di r\;r^2p(r)|_{\mbox{\footnotesize 2-derivative term}}
 = 13.7\,{\rm MeV}.
\ee
For $m_\pi\neq 0$ there is, in addition, a contribution of the mass term which
is negative for all $r$ and provides additional attraction. For the physical
pion mass the decomposition analog to (\ref{Eq:stability-in-detail-0}) reads
\be\label{Eq:stability-in-detail-138MeV}
     \int\limits_0^\infty\di r\;r^2p(r)|_i = \cases{
     \phantom{-}15.4 \,{\rm MeV} & for $i=$ 4-derivative term, \cr
               -12.3 \,{\rm MeV} & for $i=$ 2-derivative term, \cr
     -\phantom{1}3.1 \,{\rm MeV} & for $i=$ mass-term.}
\ee

%
\begingroup
\squeezetable
\begin{table}[t!]
    \caption{\footnotesize\sl
    \label{Table-II}
  Different quantities related to the nucleon EMT form factors and their
  densities:
  the energy density in the center of the nucleon $T_{00}(0)$,
  the mean square radii $\la r_E^2\ra$ and $\la r_J^2\ra$ defined in
  Eqs.~(\ref{Eq:def-energy-mean-square-radius},
  \ref{Eq:def-ang-mom-mean-square-radius}),
  the pressure $p(0)$ in the center of the nucleon,
  the position $r_0$ of the zero of the pressure defined as $p(r_0)=0$,
  the constant $d_1$,
  the dipole masses of the form factors $M_2(t)$, $J(t)$ and $d_1(t)$
  as defined in Eq.~(\ref{Eq:dipol}),
  and the mean square radius of the operator of the EMT trace
  $\la r_{\rm tr}^2\ra$.
  In the chiral limit $J(t)$ and $d_1(t)$ have infinitely steep slopes at
  $t=0$, see text, and dipole fits do not provide useful approximations.
  For sake of comparison results from the chiral quark soliton model are
  shown \cite{Goeke:2007fp}.}
\vspace{0.2cm}
    \begin{ruledtabular}
    \begin{tabular}{ccccccccrcll}
    \\
&
$m_\pi$     &
$T_{00}(0)$ &
$\la r_E^2\ra$  &
$\la r_J^2\ra$  &
$p(0)$      &
$r_0$       &
$d_1$       &
\multicolumn{3}{l}{dipole masses $M_{\rm dip}$ in GeV for:} &
$\la r_{\rm tr}^2\ra$  $\phantom{XXXX}$
\cr
&
${\rm MeV}$ &
${\rm GeV}/{\rm fm}^3$&
${\rm fm}^2$&
${\rm fm}^2$&
${\rm GeV}/{\rm fm}^3$&
${\rm fm}$ &
&
$\phantom{XX}$ $M_2(t)$ &
$J(t)$   &
$d_1(t)$ $\phantom{XX}$ &
$\;{\rm fm}^2$ \cr
    \\
    \hline
    \\
\multicolumn{5}{l}{{Skyrme model, results obtained here:}} &&&&&\cr
\cr
&0  & 1.58& 0.89& $\infty$& 0.317& 0.70& -6.60& 1.00 & --- & --- & 1.47 \\
&138& 2.28& 0.54& 0.92    & 0.477& 0.64& -4.48& 1.16 & 0.99& 0.69& 0.93 \\
\\
\multicolumn{5}{l}{{chiral quark soliton model, Ref.~\cite{Goeke:2007fp}:}} &&&&&\cr
\cr
&0  & 1.54& 0.79& $\infty$& 0.195& 0.59& -3.46& 0.87 &  ---& --- & 1.01 \\
&138& 1.70& 0.67& 1.32    & 0.232& 0.57& -2.35& 0.91 & 0.75& 0.65& 0.81 \\

\cr
\end{tabular}
\end{ruledtabular}
\end{table}
\endgroup
%

\section{\boldmath
Chiral properties of the form factors}
\label{Sec-6:ffs-chiral-properties}

Before we turn to the study of the chiral properties of the EMT form factors
(\ref{Eq:M2-d1-model-comp},~\ref{Eq:d1-model-comp},~\ref{Eq:J-model-comp})
it is important to remark that the limits $N_c\to\infty$ and $m_\pi\to 0$
do not commute \cite{Dashen:1993jt}.
The reason for that is the special role played by the $\Delta$-resonance.
The mass difference $M_\Delta-M_N={\mathcal O}(N_c^{-1})$ vanishes for $N_c\to\infty$,
and one has to consider the $\Delta$ on equal footing to
the nucleon as an intermediate state in chiral loops.

One consequence of the non-commutativity of these limits concerns
the coefficients of the leading non-analytic (in the current quark masses)
contributions (i.e.\ terms proportional to odd powers or logarithms of $m_\pi$),
which are model-independent --- i.e.\ must follow from any consistent chiral approach.
These coefficients depend on whether $N_c$ is kept finite or taken to infinity.
For example the chiral expansion of the nucleon mass reads
\be\label{Eq:Mn-mpi}
    M_N(m_\pi) = A + B\,m_\pi^2 - k_I\;\frac{3g_A^2}{8\pi F_\pi^2}\;m_\pi^3
    + \dots \;\;\;\mbox{with}\;\;\; k_I=\cases{1 & for $N_c$ finite \cr
                                             3 & for $N_c\to\infty$.}
\ee
For scalar-isoscalar quantities the ``discrepancy'' is found to be always $k_I=1$
for finite $N_c$ vs.\  $k_I=3$ for large $N_c$. For vector-isovector quantities
a different factor $k_V$ appears which is $k_V=1$ for finite $N_c$, and  $k_V=\frac32$ for
$N_c\to\infty$. The factors $k_i$ follow from soliton symmetries \cite{Cohen:1992uy}.
The form factors $M_2(t)$ and $d_1(t)$ are of scalar-isoscalar type.

Scalar-isovector or vector-isoscalar quantities ($J(t)$ is an example for the latter)
appear at subleading order of the large-$N_c$ expansion and arise in the soliton
approach from soliton rotations. For such quantities the non-commutativity of the
limits $N_c\to\infty$ and $m_\pi\to 0$ exhibits a more complicated pattern.
A first example for that was encountered in the Skyrme model, where the
isovector electric mean square radius diverges in the chiral limit as $1/m_\pi$
\cite{Adkins:1983hy} --- and not as ${\rm ln }\,m_\pi$ \cite{Beg:1973sc}.
The reason for that is that by limiting oneself to $1/N_c$-corrections due
to soliton rotations only, one ignores pion-loop corrections which appear at
the same order in $1/N_c$. The probably most advanced 
exploration of such corrections was given in \cite{Meier:1996ng}.
Taking them here into account would go beyond the scope of the 
present study.

Let us discuss first the chiral properties of the form factor $d_1(t)$.
Its chiral expansion at $t=0$ reads, see App.~\ref{App:chiral-properties},
\be\label{Eq:chi-prop-d1}
    d_1 = \kringel{d_1} + \frac{15 g_A^2 M_N}{16\pi F_\pi^2}\;m_\pi
        + {\mathcal O}(m_\pi^2)\;.
\ee
Here and in the following the superscript $\!\!\kringel{\phantom{a}}$ 
above some quantity denotes its value in the chiral limit. 
Thus, $d_1$ is finite in the chiral limit.
Noteworthy it receives a large leading non-analytic (linear in $m_\pi$)
contribution.

From the large-$r$ behaviour of the distributions of pressure and shear forces
in Eqs.~(\ref{Eq:pressure-large-r},~\ref{Eq:shear-large-r}) we concluded that
the derivative of $d_1(t)$ at zero momentum transfer diverges in the chiral
limit. More precisely, we obtain, see App.~\ref{App:chiral-properties},
\be\label{Eq:chi-prop-d1prime}
    d_1^\prime(0) = - \frac{3g_A^2M_N}{8\pi F_\pi^2 \, m_\pi}
     +{\mathcal O}(m_\pi^0)\;.
\ee

For the chiral expansion of the mean square radius of the energy density,
which is defined in (\ref{Eq:def-energy-mean-square-radius}) and which
will be needed in the discussion of the behaviour of the form factor $M_2(t)$,
we obtain, see App.~\ref{App:chiral-properties},
\be\label{Eq:chi-prop-rE2}
    \la r_E^2\ra =\la \kringel{r}{\!}_E^2\ra
    - \frac{81 g_A^2 \, m_\pi}{16\pi F_\pi^2 M_N} + {\mathcal O}(m_\pi^2)\;.
\ee
Thus, $\la r_E^2\ra$ increases in the chiral limit --- as all mean square
radii do --- but remains finite.
The normalization of the form factor $M_2(t)$ at zero momentum transfer is fixed
to unity, see Eq.~(\ref{Eq:M2-J-d1}). However, its derivative which is given
in terms of $\la r_E^2\ra$ and $d_1$, see below Eq.~(\ref{Eq:chi-prop-M2prime}),
has a non-trivial chiral expansion, namely
\be\label{Eq:chi-prop-M2prime}
    M_2^\prime(0) \equiv \frac{d_1}{5 M_N^2} + \frac{\la r^2_E\ra}{6}
    =  \;\kringel{\!\! M}{\!}_2^\prime(0)
    -  \frac{39 g_A^2 \, m_\pi}{8\pi F_\pi^2 M_N} + {\mathcal O}(m_\pi^2)\,
\ee
which follows from combining the results from
Eqs.~(\ref{Eq:chi-prop-d1},~\ref{Eq:chi-prop-rE2}).

Also the normalization of $J(t)$ at zero-momentum transfer is fixed, see
Eq.~(\ref{Eq:M2-J-d1}), and what is of interest is e.g.\  $J^\prime(0)$
whose chiral behaviour follows from that of the mean square radius of the
angular momentum density (\ref{Eq:def-ang-mom-mean-square-radius}), namely
\be\label{Eq:chi-prop-Jprime-rJ2}
    J^\prime(0) = \frac{\la r_J^2\ra}{6}\;,\;\;\;
    \la r^2_J\ra = \frac{15 g_A^2}{8\pi F_\pi^2\Theta m_\pi}
    + {\mathcal O}(m_\pi^0)\;,
\ee
see App.~\ref{App:chiral-properties}. We see that $\la r_J^2\ra$ diverges
in the chiral limit which confirms our qualitative conclusion drawn from
the large-$r$ behaviour of the density $\rho_J(r)$ in
Sec.~\ref{Sec-5:properties-of-the-densities-and-results}.
In (\ref{Eq:chi-prop-Jprime-rJ2}) one could eliminate the soliton moment of
inertia $\Theta$ in favour of the $\Delta$-nucleon mass-splitting
$M_\Delta-M_N=\frac{3}{2\Theta}$.

The results for the chiral expansions in
Eqs.~(\ref{Eq:chi-prop-d1}-\ref{Eq:chi-prop-Jprime-rJ2})
agree with the results obtained in the CQSM \cite{Goeke:2007fp}.
This is not surprizing.
Both models describe the nucleon as chiral solitons whose structure is
characterized by certain soliton profile functions. Those are determined
by different dynamics, but what they have in common is the long-distance
behaviour (\ref{Eq:prof-large-r}) uniquely fixed by chiral symmetry.
(It is worthwhile remarking that it is possible to define a formal limit 
in which the Skyrme model expressions follow from the chiral quark soliton 
model \cite{Praszalowicz:1995vi}.)

For $d_1(t)$ and $M_2(t)$ which appear in leading order in the large $N_c$ expansion
our results agree with those obtained in chiral perturbation theory at finite $N_c$
\cite{Diehl:2006ya} --- up to the expected factor $k_I=1$ vs.\ $3$,
see Eq.~(\ref{Eq:Mn-mpi}).
For $J(t)$ which appears at subleading order of the large $N_c$ expansion
we obtain $J^\prime(0)\propto 1/m_\pi$ vs. ${\rm ln}\,m_\pi$
at finite $N_c$ \cite{Diehl:2006ya} which is also expected,
see the discussion above.

In the next Section we will discuss up to which values of $m_\pi$ the chiral
expansions (\ref{Eq:chi-prop-d1}-\ref{Eq:chi-prop-Jprime-rJ2}) provide useful
approximations for the exact model results.

\section{\boldmath Results for the form factors}
\label{Sec-7:Results-ffs}

\begin{figure*}[t!]
\begin{tabular}{ccc}
\hspace{-1.0cm}  \includegraphics[height=4.8cm]{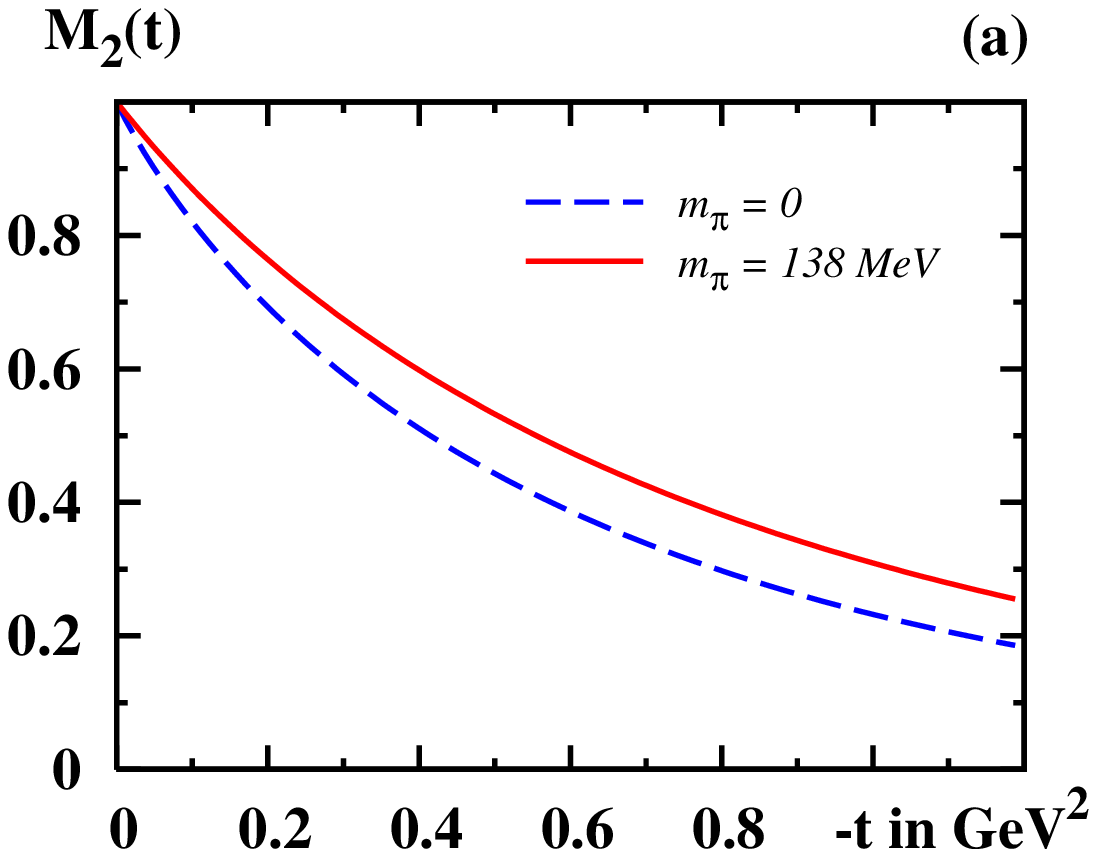}&
\hspace{-0.3cm}  \includegraphics[height=4.8cm]{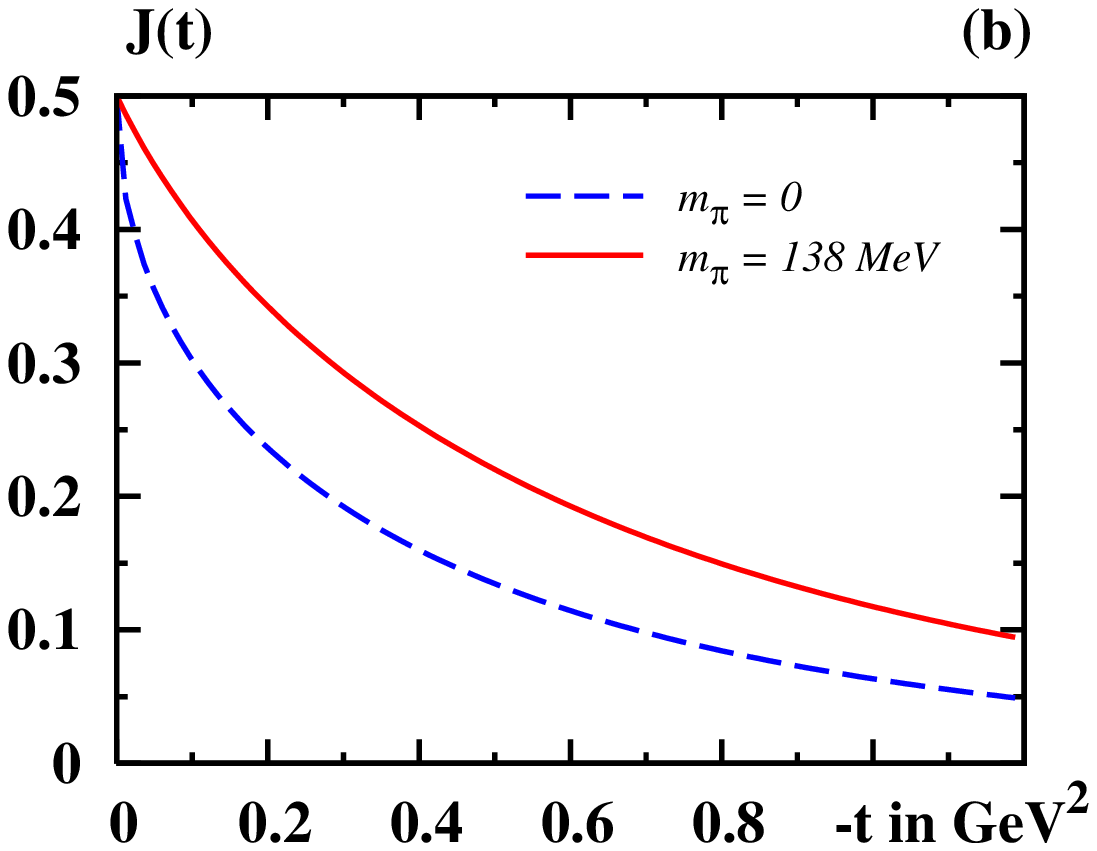}&
\hspace{-0.3cm}  \includegraphics[height=4.8cm]{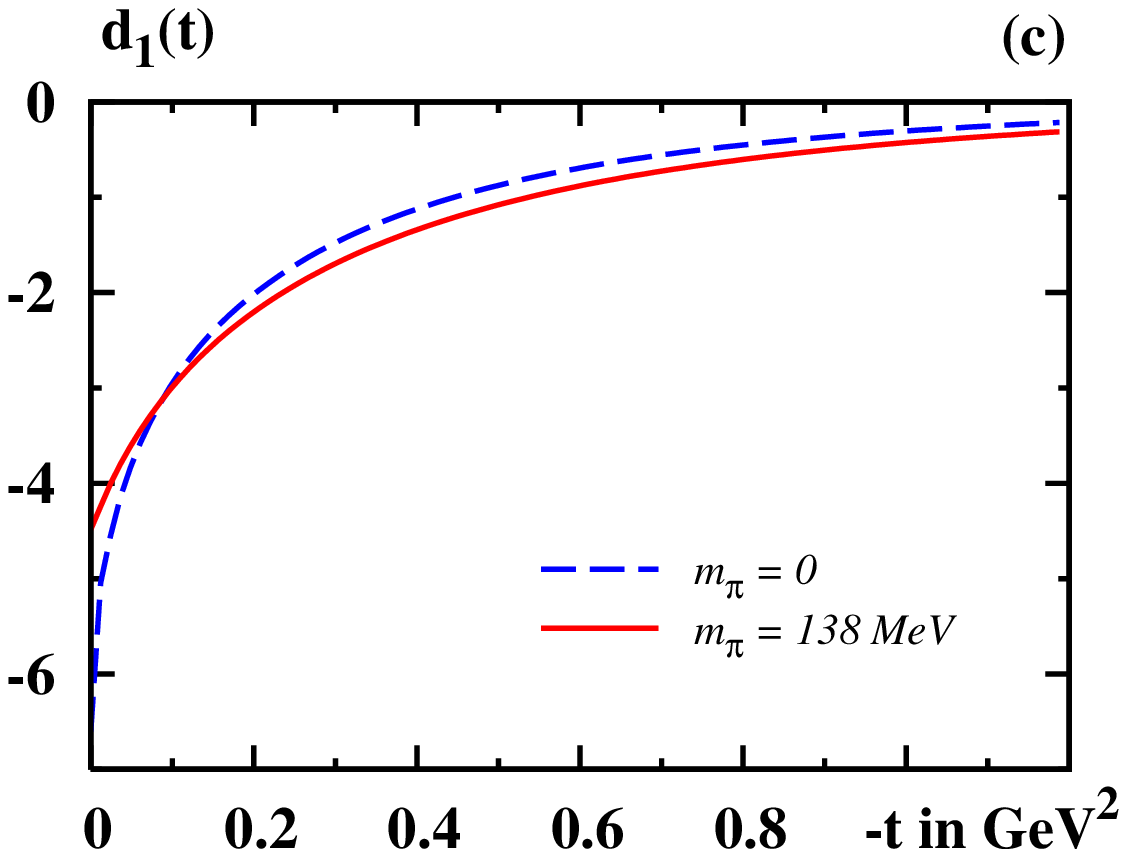}
\end{tabular}
    \caption{\label{Fig5-ffs}
    \footnotesize\sl
    The nucleon EMT form factors $M_2(t)$, $J(t)$ and $d_1(t)$
        as functions of $t$ in the Skyrme model for
    the physical value of the pion mass (solid lines)
        and in the chiral limit (dashed lines).
    In the chiral limit $J(t)$ and $d_1(t)$ exhibit infinitely steep
    slopes at $t=0$, see text.}
\begin{tabular}{cc}
  \includegraphics[height=4.8cm]{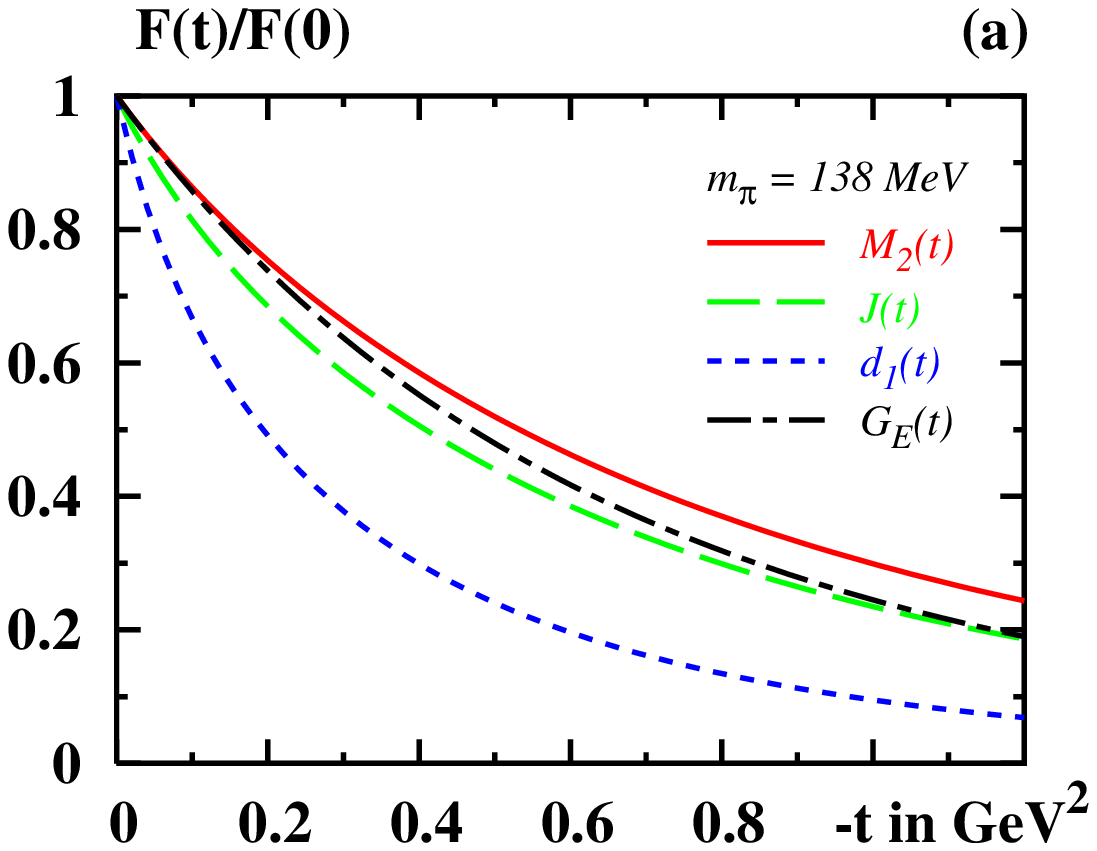}&
  \includegraphics[height=4.8cm]{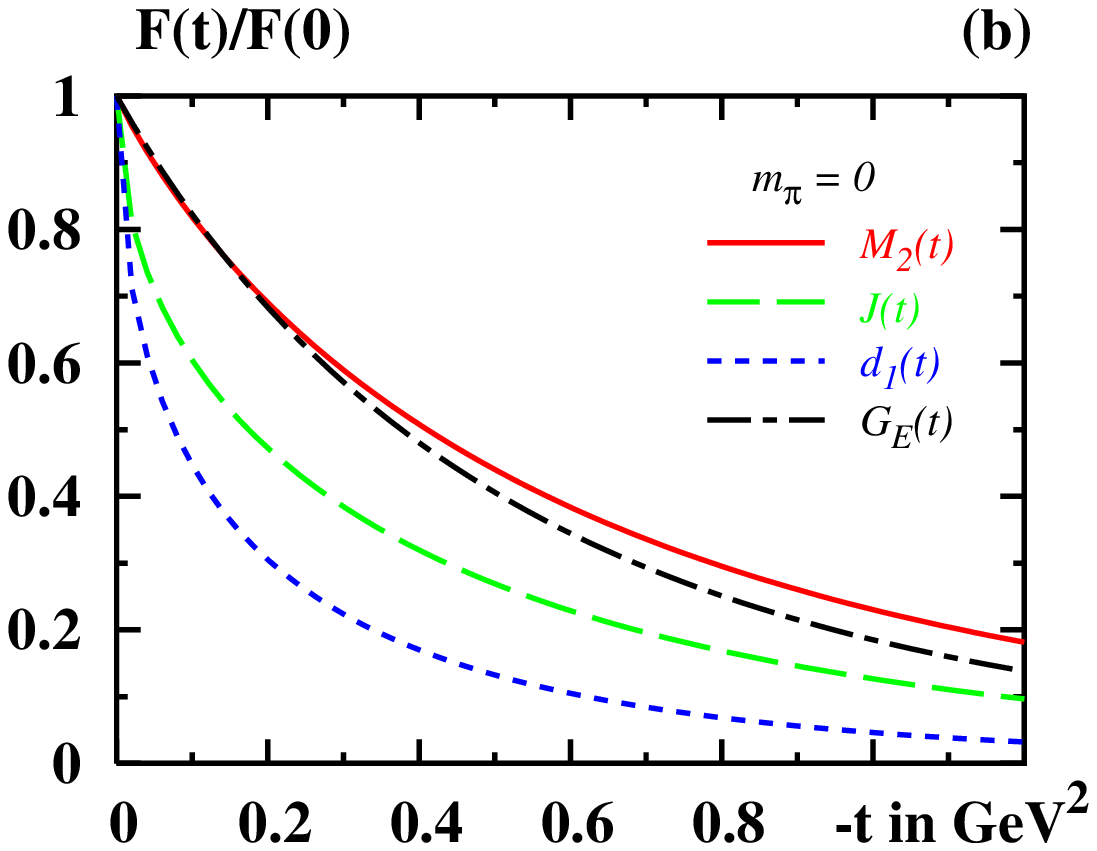}
\end{tabular}
    \caption{\label{Fig6-compare-ffs}
    \footnotesize\sl
    Nucleon form factors normalized with respect to their values at $t=0$
    in the Skyrme model as functions of $t$: The EMT form factors
    $M_2(t)$ (solid line),
    $J(t)$ (dashed line),
    $d_1(t)$ (dotted line), and the isoscalar electric form factor
    $G_E(t)$ (dashed-dotted line) for the physical value of the pion mass
    (a), and in the chiral limit (b).}
\end{figure*}

Taking according to Eqs.~(\ref{Eq:M2-d1-model-comp}-\ref{Eq:J-model-comp})
the Fourier transforms of the ``densities''  $T_{00}(r)$, $\rho_J(r)$,
$p(r)$ discussed in Sec.~\ref{Sec-5:properties-of-the-densities-and-results}
yields the nucleon EMT form factors. 
The large-$N_c$ approach can describe form factors for $|t|\ll M_N^2$. 
Here we restrict ourselves to $0\le (-t)\lesssim 1\,{\rm GeV}^2$.
In this region of $t$ relativistic corrections were estimated not to exceed 
$30\,\%$ \cite{Ji:1991ff} which is within the accuracy of the model. 
The results for the nucleon EMT form factors $M_2(t)$, $J(t)$, $d_1(t)$ are shown
in Fig.~\ref{Fig5-ffs}. For the physical pion mass all EMT form factors can be well
approximated to within an accuracy of (1-2)$\%$ by dipoles of the type
\be\label{Eq:dipol}
    F(t) \approx \frac{F(0)}{(1-t/M_{\rm dip}^2)^2}
        \;\;\;\mbox{with}\;\;\;M_{\rm dip}=\cases{
          1.17\,{\rm GeV} & for $F(t)=M_2(t)$\cr
          0.99\,{\rm GeV} & for $F(t)=J(t)$\cr
          0.69\,{\rm GeV} & for $F(t)=d_1(t)$}
\ee
for $0 \le (-t) \lesssim 0.8\,{\rm GeV}^2$.
Fig.~\ref{Fig5-ffs} shows the form factors also in the chiral limit.
Here dipole approximations are not useful for $J(t)$ and $d_1(t)$
which exhibit infinitely steep slopes at zero momentum transfer.

It is instructive to compare within the model the EMT form factors to the
electromagnetic form factors --- for example to the electric isocalar form factor
$G_E^{I=0}(t)=G_E^p(t)+G_E^n(t)\equiv G_E(t)$ \cite{Braaten:1986iw}.
Fig.~\ref{Fig6-compare-ffs} shows $G_E(t)$, $M_2(t)$, $J(t)$, $d_1(t)$
normalized with respect to their values at $t=0$ where necessary.
For the physical pion mass  $M_2(t)$ and $J(t)$ exhibit a similar $t$-dependence
as $G_E(t)$, while $d_1(t)$ exhibits a faster fall off,
see Fig.~\ref{Fig6-compare-ffs}a.
The similarity of $M_2(t)$ and $G_E(t)$ persists in the chiral limit,
but $d_1(t)$ and $J(t)$ have infinitely steep slopes at $t=0$
and fall off much faster with decreasing $t$ than $M_2(t)$ and $G_E(t)$,
see Fig.~\ref{Fig6-compare-ffs}b.

\begin{figure*}[t!]
\begin{tabular}{cc}
  \includegraphics[height=4.4cm]{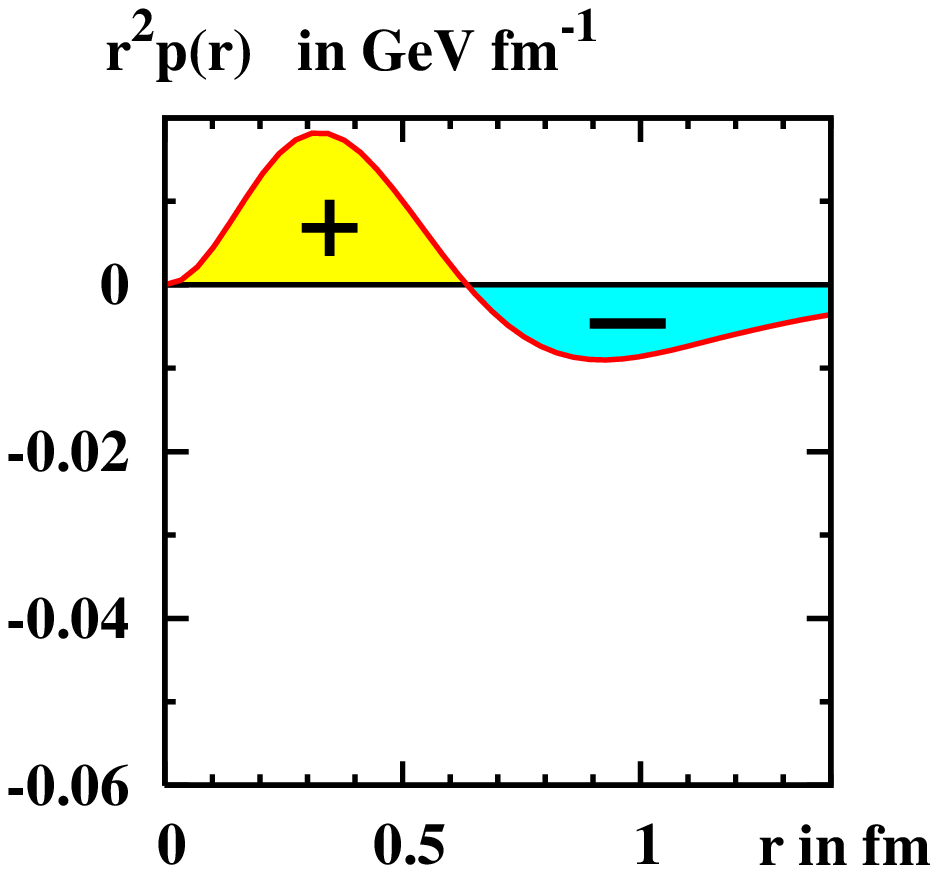}&
  \includegraphics[height=4.4cm]{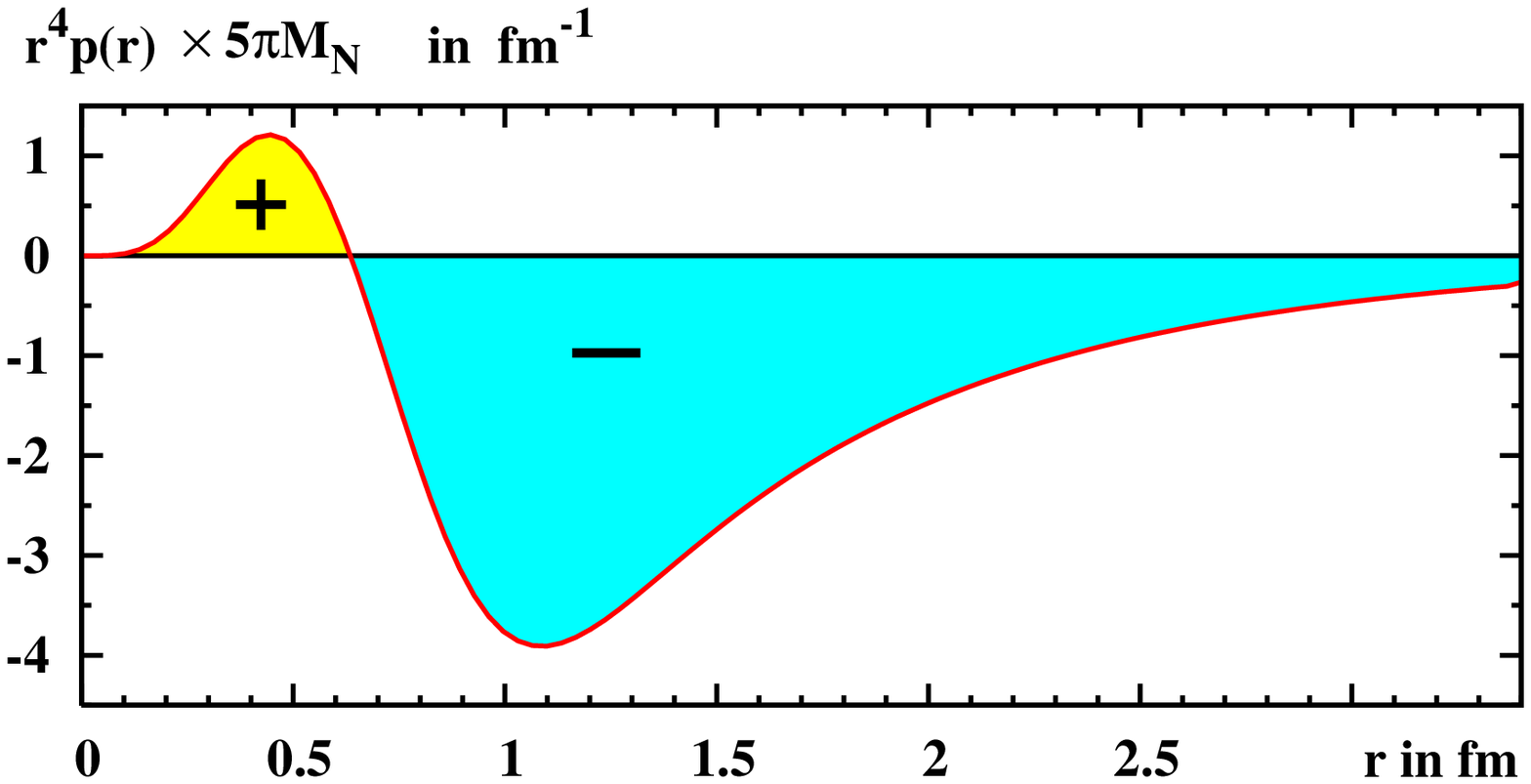}
\end{tabular}
    \caption{\label{Fig7-sign-of-d1}
    \footnotesize\sl
    Relation of stability and the sign of $d_1$.
    (a) $r^2p(r)$ as function of $r$ for the physical pion mass.
    Integrating the curve over $r$ yields zero because the
    inner-positive-pressure-region precisely cancells the
    outer-negative-pressure-region.
    (b) $r^4p(r)$ multiplied by the constant $\frac54 M_N 4\pi$
    as function of $r$. Integrating this curve over $r$
    yields $d_1$ which apparently has a negative sign.}
\end{figure*}

No principle fixes the normalization of the form factor $d_1(t)$ at
zero-momentum transfer --- in contrast to $M_2(t)$ and $J(t)$.
On the basis of stability requirements it was conjectured that
$d_1 \equiv d_1(0)$ must be negative \cite{Goeke:2007fp}, i.e.\
\be\label{Eq:d1-sign}
    d_1 <0 \;.
\ee
In fact, the conservation of the EMT dictates $\int_0^\infty\di r\, r^2p(r)=0$.
For a mechanically stable object the pressure must be positive (repulsion) in
the inner region $r<r_0$, and negative (attraction) in the outer region $r>r_0$,
as we observe also in the Skyrme model, see
Sec.~\ref{Sec-5:properties-of-the-densities-and-results}.
This picture of the distribution of pressure forces, Fig.~\ref{Fig7-sign-of-d1}a,
immediately dictates a negative sign for $d_1\propto\int_0^\infty\di r\,r^4p(r)=0$,
Fig.~\ref{Fig7-sign-of-d1}b.
Thus, the Skyrme model fully confirms the conclusion (\ref{Eq:d1-sign}).

Numerically we obtain $d_1=-4.48$ at the physical value of the pion mass 
for our choice of parameters, which confirms not only the sign but also 
the magnitude of results from the chiral quark soliton model
\cite{Wakamatsu:2006dy,Goeke:2007fp,Goeke:2007fq,Petrov:1998kf,Kivel:2000fg,Schweitzer:2002nm}. 

\begin{figure*}[t!]
\begin{tabular}{cccc}
        \includegraphics[height=4cm]{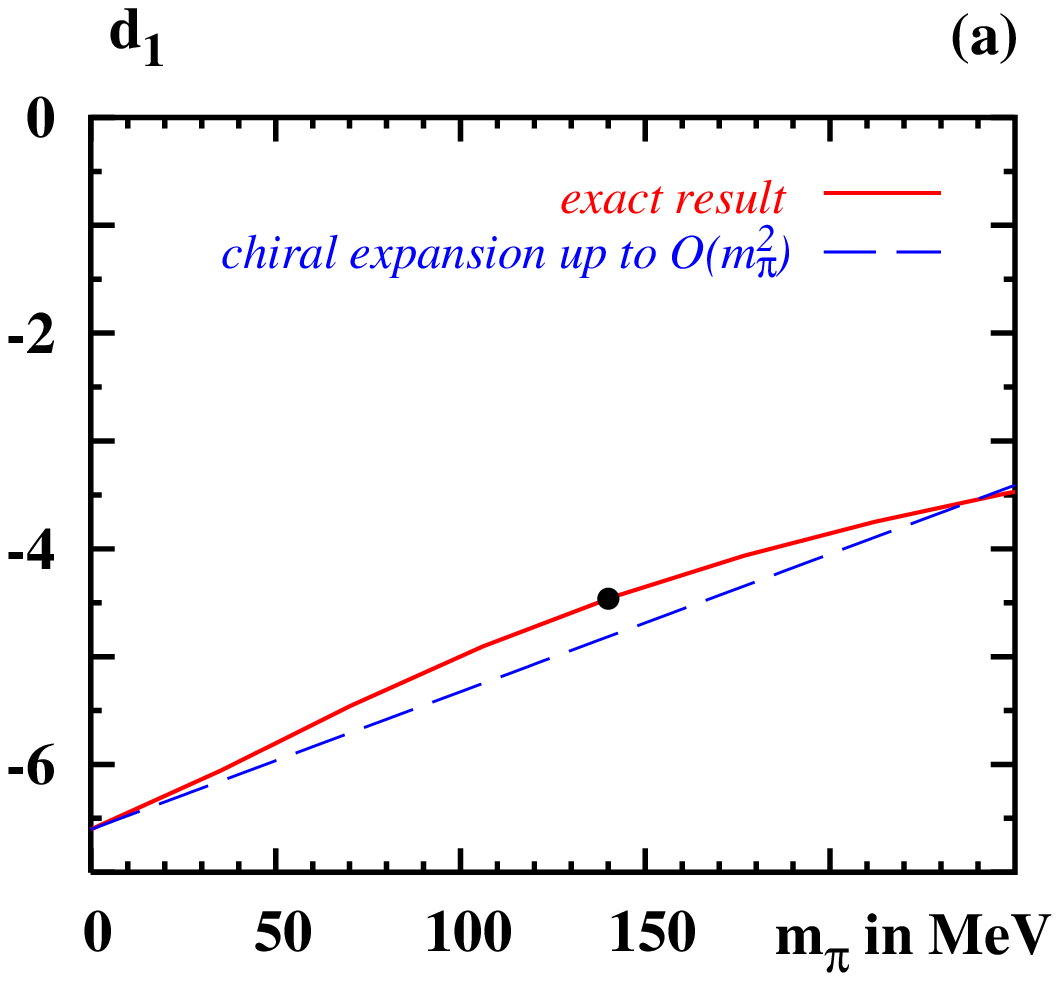} &
    \includegraphics[height=4cm]{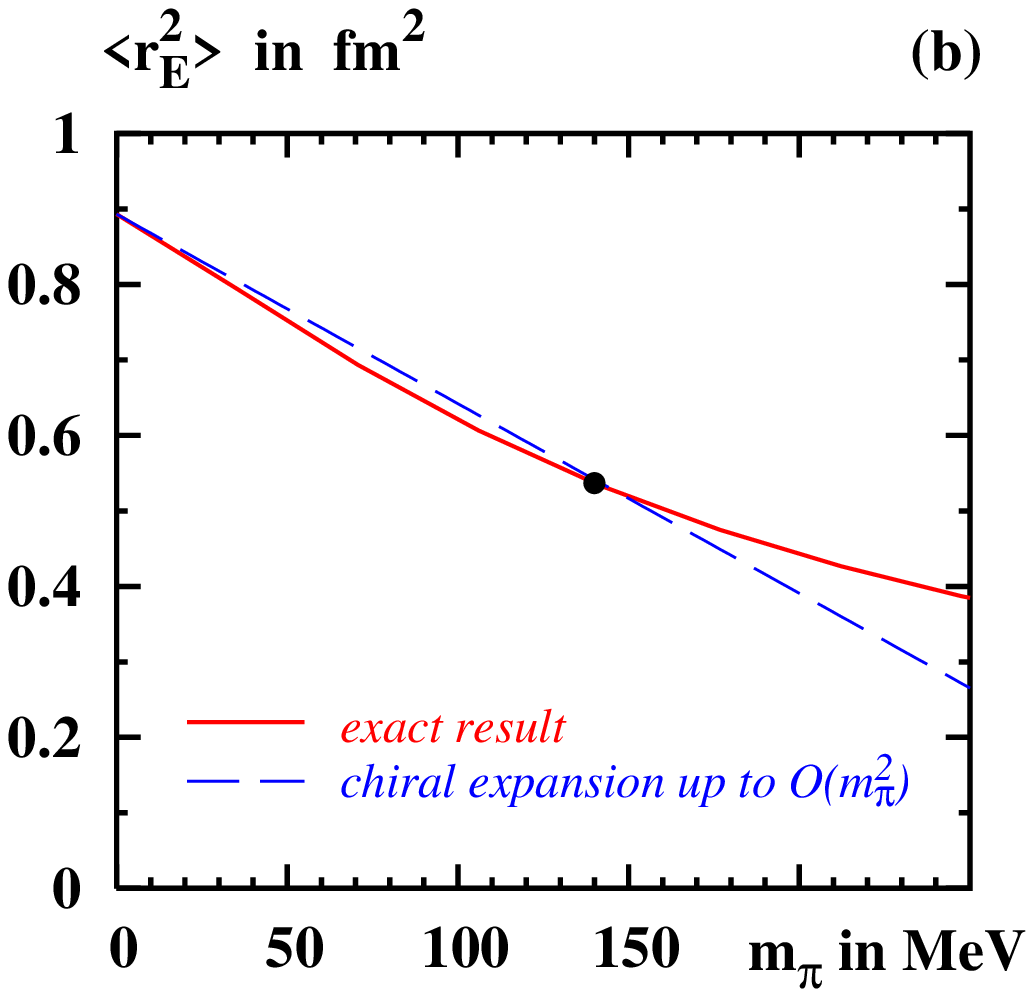} &
        \includegraphics[height=4cm]{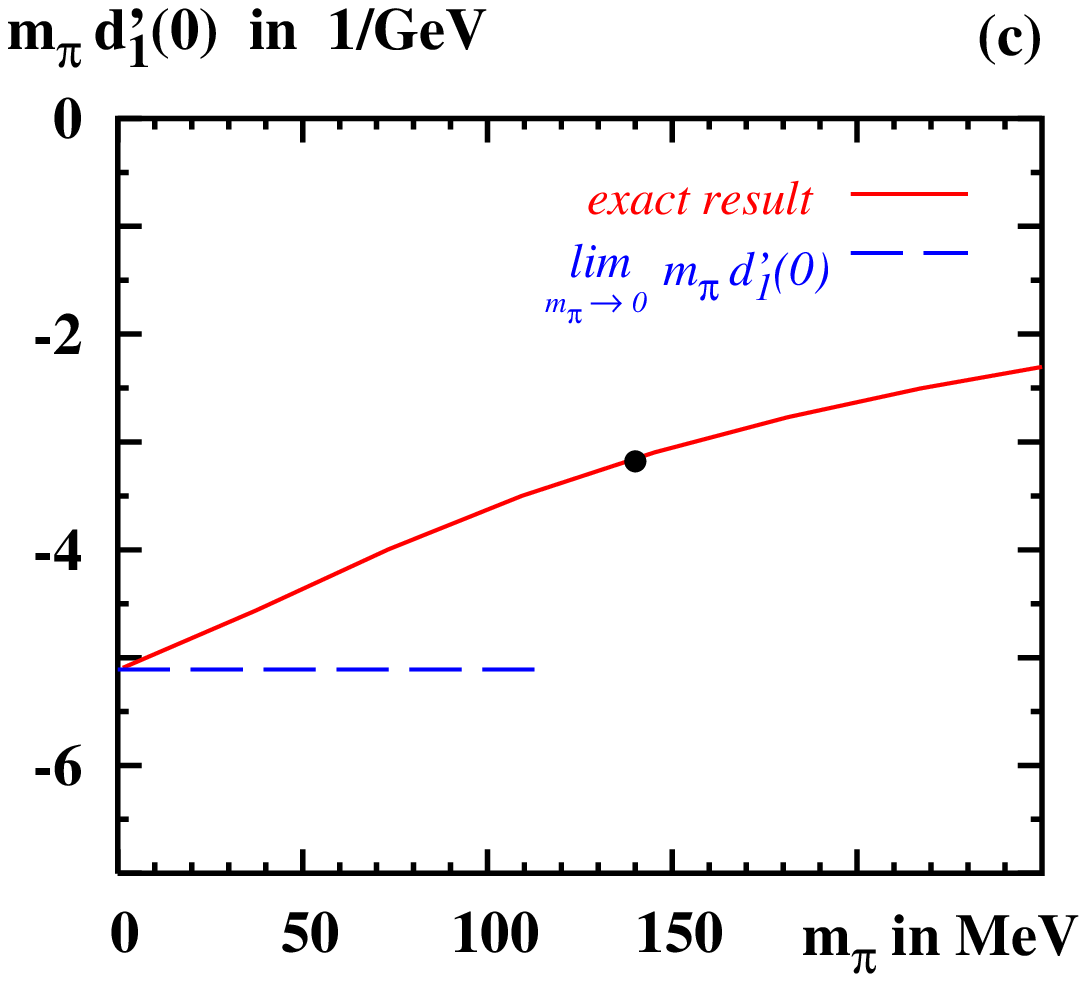} &
    \includegraphics[height=4cm]{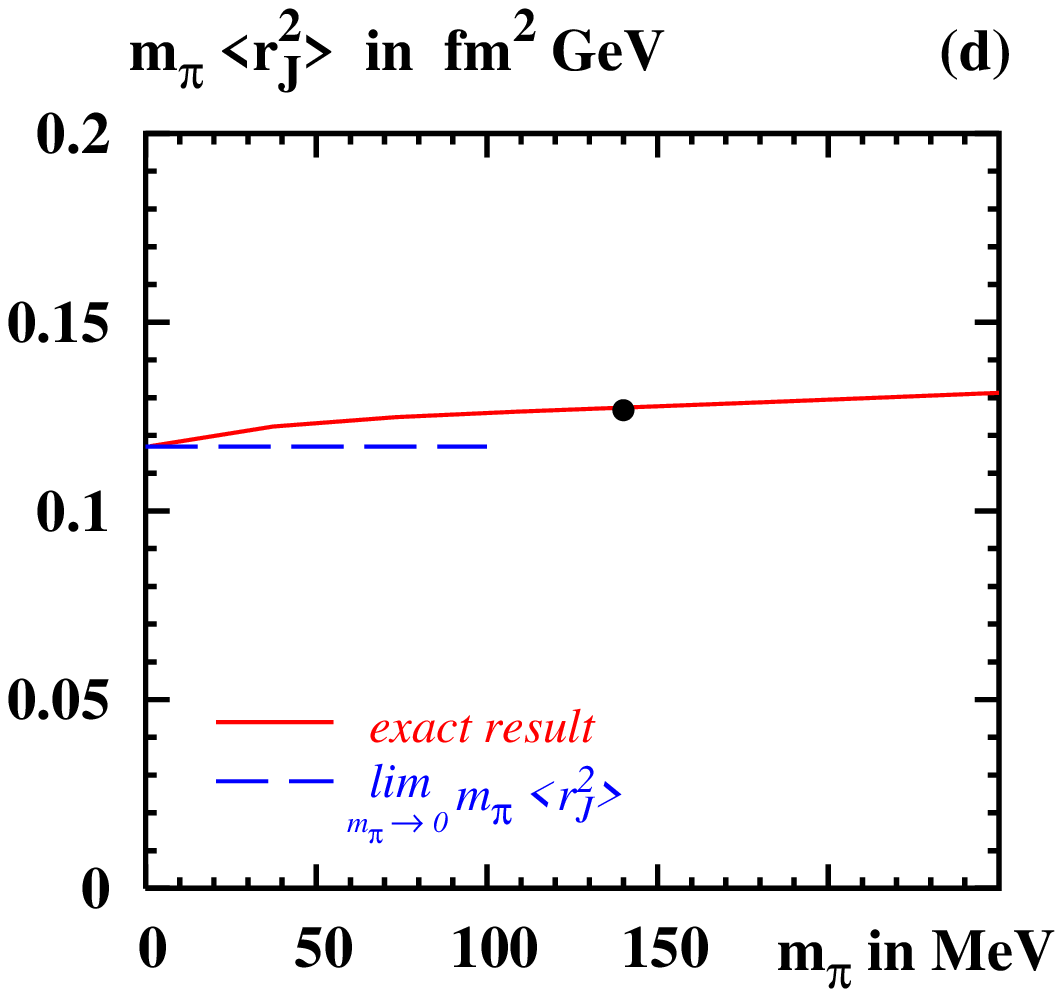}
\end{tabular}
    \caption{\label{Fig8-chiral properties}
    \footnotesize\sl
    (a+b): The constant $d_1$ and the mean square radius of the energy density
    $\la r_E^2\ra$ as functions of $m_\pi$. Solid lines: exact results.
    Dashed lines: Chiral expansions up to ${\mathcal O}(m_\pi^2)$ according to
    Eqs.~(\ref{Eq:chi-prop-d1},~\ref{Eq:chi-prop-rE2}).
    (c+d): The numerical confirmations that $d_1^\prime(0)$ and 
    $\la r_J^2\ra$ diverge in the chiral limit as $1/m_\pi$ according to
    Eqs.~(\ref{Eq:chi-prop-d1prime},~\ref{Eq:chi-prop-Jprime-rJ2}). }
\end{figure*}

It is also of interest to compare the chiral expansions in
Eqs.~(\ref{Eq:chi-prop-d1}-\ref{Eq:chi-prop-Jprime-rJ2}) to exact model results
which is done in Fig.~\ref{Fig8-chiral properties}.
Figs.~\ref{Fig8-chiral properties}a, b show that the exact chiral behaviour
of $d_1$ and the mean square radius of the energy density in the model is
dominated by the leading non-analytic contributions in
Eqs.~(\ref{Eq:chi-prop-d1},~\ref{Eq:chi-prop-rE2}) up to the physical value of
the pion mass.
Figs.~\ref{Fig8-chiral properties}c, d show that the slope of $d_1(t)$ at $t=0$
and the mean square radius of the angular momentum density diverge in the
chiral limit as $1/m_\pi$ precisely as predicted by
Eqs.~(\ref{Eq:chi-prop-d1prime},~\ref{Eq:chi-prop-Jprime-rJ2}).
(The smallest finite value of $m_\pi$ computed numerically
in Figs.~\ref{Fig8-chiral properties}c, is about $1\,{\rm MeV}$.)

Finally, we turn our attention to the trace of the EMT. Due to the trace
anomaly  in QCD \cite{Adler:1976zt} the operator of the trace of the total
EMT is given by
\be\label{Eq:ff-of-EMT-trace-1}
    \hat{T}_{\!\mu}^{\,\mu} \equiv \frac{\beta}{2g}\;F^{\mu\nu}F_{\mu\nu}
    +(1+\gamma_m)\sum_am_a\bar\psi_a\psi_a \;.\ee
Let $F_{\rm tr}(t)$ denote the scalar form factor of the operator
(\ref{Eq:ff-of-EMT-trace-1}) which is, of course, not independent but a
linear combination of the EMT form factors.
The small-$t$ expansion of $F_{\rm tr}(t)=1+t\la r^2_{\rm tr}\ra/6+{\mathcal O}(t^2)$
defines the mean square radius $\la r_{\rm tr}^2\ra$ which is related to
$\la r_E^2\ra$ and $d_1$ as \cite{Goeke:2007fp}
\be\label{Eq:ff-of-EMT-trace-2}
        \la r_{\rm tr}^2\ra =
        \frac{\int\di^3{\bf r}\,r^2T_\mu^\mu(r)}{\int\di^3{\bf r}\,T_\mu^\mu(r)}
        = \la r_E^2\ra -\frac{12\,d_1}{5M_N^2}
    = \la \kringel{r}{\!}_{\rm tr}^2\ra
    - \frac{117 g_A^2}{16\pi F_\pi^2 M_N}\;m_\pi
        + {\mathcal O}(m_\pi^2)\;,
\ee
where the chiral expansion follows from
Eqs.~(\ref{Eq:chi-prop-d1},~\ref{Eq:chi-prop-rE2}).
Numerically we obtain
\be\label{Eq:chi-prop-r2trace-value}
       \la r_{\rm tr}^2\ra = \cases{
    1.47\,{\rm fm}^2 & in the chiral limit, \cr
    0.93\,{\rm fm}^2 & for $m_\pi=138\,{\rm MeV}$.}
\ee
In the chiral limit $\la r_{\rm tr}^2\ra$ is the mean square radius of the
gluonic operator $F^{\mu\nu}F_{\mu\nu}$. The large value obtained in
(\ref{Eq:chi-prop-r2trace-value}) is rather interesting, in particular,
if we confront it with the mean square radius of the traceless part of the EMT
estimated by means of QCD sum rules to be about $0.1\,{\rm fm}^2$ \cite{Braun:1992jp}.
A possible explanation why the radii of the trace and the traceless part of
the EMT differ so much could be provided in the instanton vacuum model
\cite{Diakonov:1995qy}, see \cite{Goeke:2007fp} for details.

\section{Conclusions}
\label{Sec-8:conclusions}

In this work we presented a study of the 
form factors of the energy momentum tensor of the nucleon in the Skyrme
model. We have shown the theoretical consistency of the approach and provided
explicit proofs that the EMT form factors in the Skyrme model satisfy all
general requirements, provided one evaluates the respective expressions with that 
profile function which minimizes the actual expression for the nucleon mass.

We derived the chiral expansions of the form factors in the Skyrme model 
and found them in agreement with results from chiral perturbation theory 
\cite{Belitsky:2002jp,Diehl:2006ya} considering the non-commutativity of 
the limits $m_\pi\to0$ and $N_c\to\infty$ \cite{Dashen:1993jt,Cohen:1992uy}.
The Skyrme model also confirms the strong $m_\pi$-dependence of the constant $d_1$
observed in the chiral quark soliton model \cite{Goeke:2007fp,Goeke:2007fq}.
This is of practical interest in the context of extrapolating lattice QCD data 
to the physical value of the pion mass, see \cite{Goeke:2007fq}.

The numerical results for the form factors are in good qualitative agreement 
with results from the chiral quark soliton model \cite{Goeke:2007fp}. 
Also the Skyrme model yields a negative constant $d_1$ and confirms the 
conclusions drawn in \cite{Goeke:2007fp} that $d_1$ must be negative for 
a mechanically stable object. Numerically we find for the constant 
$d_1=-4.88$ in agreement with estimates from the chiral quark soliton model
\cite{Wakamatsu:2006dy,Goeke:2007fp,Goeke:2007fq,Petrov:1998kf,Kivel:2000fg,Schweitzer:2002nm}. 

Results from both models suggest that all form factors can be well approximated 
by dipole fits for $0 \le (-t) \lesssim 1\,{\rm GeV}^2$. $M_2(t)$ and $J(t)$ 
exhibit a similar $t$-dependence as the isoscalar electric form factor, while 
$d_1(t)$ shows a much faster fall-off with decreasing $t$. Both the Skyrme and 
chiral quark soliton model \cite{Goeke:2007fp} predict for $d_1(t)$ a dipole mass 
of $0.7\,{\rm GeV}$.
The predictions for the $t$-dependence of $d_1(t)$ could contribute to an 
explanation of the $t$-dependence of the beam-charge asymmetry observed at 
HERMES \cite{Airapetian:2006zr}.
Of particular interest is also the predicted $t$-dependence of the form factor 
$J(t)$ which is of relevance for extracting from data the quark angular momentum 
contribution to the nucleon spin \cite{Ellinghaus:2005uc}.
From the experience with the description of other observables in the
Skyrme model we expect our predictions to hold to within an accuracy of 
(30--40)$\%$ though, of course, as always data will have the final say.

In this work we used the simplest version of the Skyrme model \cite{Skyrme:1961vq}.
It would be interesting to study the nucleon EMT form factors in extended
versions of the Skyrme model which include e.g.\  vector-mesons 
\cite{Adkins:1983nw,Meissner:1987ge,Schwesinger:1988af},
in order to see how these degrees of freedom contribute to the stability 
of the soliton and to the constant $d_1$.

\vspace{0.3cm}
\noindent
{\bf Acknowledgements} \hspace{0.2cm}
We thank Pavel Pobylitsa and Maxim Polyakov for fruitful discussions and valuable
comments. 
This research is part of the EU integrated infrastructure initiative
hadron physics project under contract number RII3-CT-2004-506078,
the Transregio/SFB Bonn-Bochum-Giessen, and the COSY-J\"ulich project,
and partially supported by the Graduierten-Kolleg Bochum-Dortmund
and Verbundforschung of BMBF.

\appendix
\section{\boldmath
Alternative definition of form factors}
\label{App:Alternative-definition}

By means of the Gordon identity $2M_N\bar u^\prime\gamma^\alpha u=$
$\bar u^\prime(i\sigma^{\alpha\kappa}\Delta_\kappa+2P^\alpha)u$ one can rewrite
(\ref{Eq:ff-of-EMT}) as (see e.g.\ in Ref.~\cite{Ji:1996ek})
\ba
    \la p^\prime| \hat T_{\mu\nu}^{Q,G}(0) |p\rangle
    &=& \bar u(p^\prime)\biggl[
    A^{Q,G}(t)\,\frac{\gamma_\mu P_\nu+\gamma_\nu P_\mu}{2}+
    B^{Q,G}(t)\,\frac{i(P_{\mu}\sigma_{\nu\rho}+P_{\nu}\sigma_{\mu\rho})
    \Delta^\rho}{4M_N}
    \nonumber \\
    &+& C^{Q,G}(t)\,\frac{\Delta_\mu\Delta_\nu-g_{\mu\nu}\Delta^2}{M_N}
    \pm \bar c(t)g_{\mu\nu} \biggr]u(p)\, ,
    \label{Eq-app:ff-of-EMT-alternative} \ea
where
$A^a(t)       = M_2^a(t)$, 
$A^a(t)+B^a(t)= 2\,J^a(t)$,
$C^a(t)       = \frac15\,d_1^a(t)$ with $a=Q,\,G$.
In this notation the constraints (\ref{Eq:M2-J-d1}) read $A^Q(0)+A^G(0)=1$
and $B^Q(0)+B^G(0)=0$ meaning that the total nucleon gravitomagnetic moment
vanishes.

\section{\boldmath
Alternative proof of the stability condition}
\label{App-stabilty}

The stability condition (\ref{Eq:stability}) can be proven alternatively
in the following way. Let $F(r)$ denote the profile function which minimizes
the soliton energy (\ref{Eq:Esol}), i.e.\ which satisfies (\ref{Eq:diff-eq}),
and substitute $r\to r^\prime = \lambda r$ in (\ref{Eq:Esol}). This yields
\be\label{Eq:Esol-lambda}
    M_{\rm sol}[\lambda] = \lambda^{-1} E_2 + \lambda E_4 + \lambda^{-3}E_m
\ee
where $E_2$, $E_4$ and $E_m$ are the contributions of respectively the 2-derivative,
4-derivative and mass term in  (\ref{Eq:Esol}). $M_{\rm sol}[\lambda]$ in 
(\ref{Eq:Esol-lambda}) has a minimum at $\lambda=1$, i.e.\ we have the 
conditions
\ba\label{Eq:Esol-lambda-2}
    \frac{\partial M_{\rm sol}[\lambda]}{\partial \lambda}\biggl|_{\lambda=1}
    \; &=& - E_2 + E_4 - 3 E_m \stackrel{!}{=}0 \;, \\
    \label{Eq:Esol-lambda-3}
    \frac{\partial^2M_{\rm sol}[\lambda]}{\partial\lambda^2}\biggl|_{\lambda=1}
     &=&  2 E_2 + 12 E_m > 0 \;,
\ea
which holds since the $E_i$ ($i=2,\,4,\,m$) are positive. The above considerations 
are well known, the relation (\ref{Eq:Esol-lambda-2}) is sometimes referred to as
the ``virial theorem'' in the Skyrme model \cite{Zahed:1986qz,Holzwarth:1985rb}.
For our purposes it is sufficient 
to observe that the virial theorem is nothing but the stability condition
\be\label{Eq:Esol-lambda-4}
     \int\limits_0^\infty\di r\;r^2p(r) = \frac{1}{12\pi}
     \frac{\partial M_{\rm sol}[\lambda]}{\partial \lambda}\biggl|_{\lambda=1}
     = 0\;.
\ee

\section{\boldmath
Rotational corrections for $T_{00}(r)$, $p(r)$ and $s(r)$}
\label{App:rot-corr}

The expression for a quantity in the Skyrme model is in general of the type
(\ref{Eq:general-observable}). 
In this work we have chosen to consider rotational corrections to a quantity
if and only if it receives no contribution from the leading order of the 
large-$N_c$ expansion. This Appendix aims at making the arguments in favour 
of this procedure more clear.

In our approach we have, in particular, neglected  rotational corrections 
to the densities $T_{00}(r)$, $p(r)$ and $s(r)$. What would have happened 
if one kept these rotational corrections?
Let us denote the densities with {\sl included} rotational corrections
by $\rotcorr{T}_{00}(r)$, $\rotcorr{p}(r)$ and $\rotcorr{s}(r)$.
The corresponding model expressions read
\ba \label{App:T00+rot}
    \rotcorr{T}_{00}(r) &=& T_{00}(r) + \frac{S(S+1)}{12\Theta^2}\;\sin^2F(r)
    \biggl[F_\pi^2+\frac{4F^\prime(r)^2}{e^2}
    +\frac{4\sin^2F(r)}{e^2r^2}\biggr]
    \;\equiv\; T_{00}(r) + \frac{S(S+1)}{\Theta}\,\rho_J(r)
    \\
    \label{App:pressure+rot}
    \rotcorr{p}(r)      &=& \; p(r)  \;+\;  \frac{S(S+1)}{12\Theta^2}\;\sin^2F(r)\,
    \biggl[F_\pi^2+\frac{4F^\prime(r)^2}{3e^2}+\frac{4\sin^2F(r)}{3e^2r^2}\biggr] 
    \\
    \label{App:shear-rot}
    \rotcorr{s}(r)      &=& \; s(r)  \;+\;\frac{S(S+1)}{12\Theta^2}\;
    \sin^2F(r)\;\biggl[\phantom{F_\pi^2}-
	\frac{8F^\prime(r)^2}{e^2}+\frac{4\sin^2F(r)}{e^2r^2}\biggr]
\ea

%
\begin{wrapfigure}[18]{R!}{7.2cm}
\vspace{-1.1cm}
\begin{center}
\includegraphics[width=6cm]{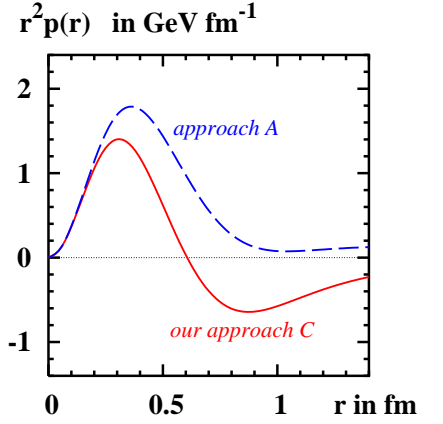}
\end{center}
\vspace{-0.6cm}
	\caption{\label{Fig9-pressure-rot-corr}
	\footnotesize\sl
	$r^2p(r)$ in the Skyrme model as function of $r$.
        Dashed line: Result obtained following the procedure (A), see text,
	used e.g.\ in Ref.~\cite{Adkins:1983ya}. 
	Solid line: Result obtained following the procedure (C) used in this
	work. Both curves refer to the chiral limit.}
\end{wrapfigure}
%
\noindent
with $T_{00}(r)$, $p(r)$ and $s(r)$ as given by
Eqs.~(\ref{Eq:T00-Skyrme},~\ref{Eq:pressure},~\ref{Eq:shear}),
i.e.\  the corresponding densities without rotational corrections. Note that 
$\int\di^3{\bf x}\,\rotcorr{T}_{00}(r)=M_{\rm sol}+\frac{S(S+1)}{2\Theta}$
with $M_{\rm sol}$ defined in (\ref{Eq:Esol}).
At this point several procedures have been considered in literature.

\vspace{0.5cm}

A. \hspace{0.1cm}
One procedure, used e.g.\  in \cite{Adkins:1983ya,Adkins:1983hy},
consists in minimizing only the soliton energy $M_{\rm sol}$
in Eq.~(\ref{Eq:Esol}), but keeping the respective rotational corrections
understood as small perturbations. Then e.g.\  the nucleon or $\Delta$ mass 
are obtained by adding up the minimized soliton energy $M_{\rm sol}$ 
and the rotational energy $\frac{S(S+1)}{2\Theta}$.
This procedure strictly speaking does not yield a stable soliton.
This can be demonstrated e.g.\  by evaluating the pressure $\rotcorr{p}(r)$
with the profile function which minimizes $M_{\rm sol}$.
For the parameters used in \cite{Adkins:1983ya}
we obtain the result shown in Fig.~\ref{Fig9-pressure-rot-corr}.
Clearly, the $\rotcorr{p}(r)$ evaluated in this way is always positive
and violates the stability condition (\ref{Eq:stability}).
This is not surprizing because from the proof of the stability
condition in the Skyrme model given in App.~\ref{App-stabilty} 
it is clear that (\ref{Eq:stability}) will be satisfied if and only if one 
evaluates the pressure with that profile function which minimizes the full energy.

\vspace{0.5cm}

B. \hspace{0.1cm}
One way to resolve the problems encountered above consists in minimizing 
$\rotcorr{M}_{\rm sol}\equiv\int\di^3{\bf x}\,\rotcorr{T}_{00}(r)$. 
The profile function which does that has to satisfy the integro-differential 
equation (since $\Theta=\Theta[F]$, see Eq.~(\ref{Eq:mom-ienertia}))
\ba\label{App:EOM-rot}
    \biggl(\mbox{left-hand side of Eq.~(\ref{Eq:diff-eq})}\biggr) + 
    \frac{S(S+1)}{6\Theta^2}\;\biggl[
    -\frac{4r^2\sin^2F(r)\,F^{\prime\prime}(r)}{e^2F_\pi^2}
    -\frac{2r^2\sin2F(r)\,F^\prime(r)^2}{e^2F_\pi^2}\;\;\;\;\;\; && \\
    -\frac{8r\sin^2F(r)\,F^\prime(r)}{e^2F_\pi^2} 
    +\frac{r^2}{2}\,\sin2F(r)+\frac{4\sin^2F(r)\,\sin2F(r)}{e^2F_\pi^2}\biggr]
    &=& 0 \nonumber
\ea
with the same boundary conditions as for Eq.~(\ref{Eq:diff-eq}).
Now one obtains a consistent description of the densities. E.g.\ 
the distributions of pressure and shear forces {\sl with} rotational 
corrections satisfy the relation (\ref{Eq:relation-p(r)-s(r)}), because 
\be\label{App:relation-p(r)-s(r)-corr}
    \frac23\;\frac{\partial\rotcorr{s}(r)}{\partial r\;}+
    \frac{2\rotcorr{s}(r)}{r} + \frac{\partial\rotcorr{p}(r)}{\partial r\;} 
    =  -\frac{F_\pi^2}{r^2}\,F^\prime(r)
    \times\biggl(\mbox{equations of motion {\sl with} rotational corrections}\biggr) 
    = 0\,, \ee
where ``equations of motion {\sl with} rotational corrections'' denotes 
the left-hand-side of Eq.~(\ref{App:EOM-rot}). Also the stability condition 
(\ref{Eq:stability}) holds with {\sl included} rotational corrections 
which can be seen by rewriting 
\ba\label{App:stability-rot}
    r^2\rotcorr{p}(r) &=& 
    \frac{\partial\;}{\partial r}\biggl\{r^3\rotcorr{p}(r) 
    +\frac{F_\pi^2}{12}\,r^3F^\prime(r)^2
    -\frac{\sin^4F(r)}{3e^2r}
    +\frac{F_\pi^2m_\pi^2}{6}\,r^3(1-\cos F(r))\nonumber\\
    && \hspace{0.5cm}
    -\frac{S(S+1)}{6\Theta^2}\biggl[ \frac{F_\pi^2}{3}r^3\sin^2F(r)
    +\frac{4r^3}{3e^2}\sin^2F(r)\,F^\prime(r)^2\biggr]    \biggr\} \nonumber\\
    &&
    -\frac{F_\pi^2}{3}\,rF^\prime(r)
    \times\biggl(\mbox{equations of motion {\sl with} rotational corrections}\biggr)\;.
\ea
Alternatively, one may use the scaling trick discussed in App.~\ref{App-stabilty}.
For that we notice that in addition to the $E_i$ in Eq.~(\ref{Eq:Esol-lambda}) 
there are two contributions due to rotational corrections, namely 
\be
	E_{\rm rot,1}=\frac{S(S+1)}{12\Theta^2}\;F_\pi^2
	\int\di^3{\bf x}\,\sin^2F(r)\;,\;\;\;\;
	E_{\rm rot,2}=\frac{S(S+1)}{3e^2\Theta^2}\int\di^3{\bf x}\,\sin^2F(r)
	\biggl(F^\prime(r)^2+\frac{\sin^2F(r)}{r^2}\biggr)
	\label{App:Erot12}
\ee
such that
\be\label{Eq:Esol-lambda-rot}
    \rotcorr{M}_{\rm sol}[\lambda] = 
	\lambda^{-1} E_2 + \lambda E_4 + \lambda^{-3}E_m
	+\lambda^{-3} E_{\rm rot,1} + \lambda^{-1} E_{\rm rot,2}\;.
\ee
Also in this case the conditions analog to (\ref{Eq:Esol-lambda-2}) hold and, 
in particular, we have
\be\label{Eq:Esol-lambda-4-rot}
     \int\limits_0^\infty\di r\;r^2\rotcorr{p}(r) = \frac{1}{12\pi}
     \frac{\partial\rotcorr{M}_{\rm sol}[\lambda]}{\partial \lambda}\biggl|_{\lambda=1}
     = 0\;.
\ee
Thus,  also with {\sl included} rotational corrections one obtains a consistent
description of the form factors of the EMT. The problem is, however, that
(\ref{App:EOM-rot}) has no solution in the chiral limit, while the solutions 
which exist for sufficiently large $m_\pi$ yield results in conflict with 
chiral symmetry \cite{Bander:1984gr,Braaten:1984qe}, 
see also the review \cite{Holzwarth:1985rb}.
 
The minimization of $\rotcorr{M}_{\rm sol}=M_{\rm sol}+\frac{S(S+1)}{2\Theta}$
would be unsatisfactory also from the point of view of large-$N_c$ counting, as 
no contribution of ${\mathcal O}(N_c^0)$ is included. Such terms could in principle
be taken into account by adopting the Peierls-Yoccoz projection technique in 
non-relativistic many body theory \cite{Birse:1984js,Fiolhais:1988ee,Neuber:1993ah}.
If one assumes the ''small overlap approximation'' the
energy is similar to the one obtained approximately by pion
corrections to the Skyrme model \cite{Zahed:1986va}. In the chiral
quark soliton model, which is in this respect similar to the
Skyrme model, this has been discussed in Ref.~\cite{Pobylitsa:1992bk}.

\vspace{0.5cm}

C. \hspace{0.1cm}
A consistent description of the nucleon EMT which respects chiral symmetry 
is obtained by adopting the procedure used in this work. 
After quantizing the soliton one neglects rotational corrections 
with respect to the leading order contribution unless the latter is zero. 
I.e.\ rotational corrections are considered if and only if the leading order 
contribution happens to vanish for symmetry reasons.

In this way one strictly speaking cannot describe the proton and neutron
separately, but has to limit oneself to isoscalar and isovector quantities 
which are in general of different order in the large-$N_c$ expansion.
This is what we did in this work, see e.g.\  Table.~\ref{Table-I}. 
However, the results obtained here for the total EMT form factors refer both to the
proton and neutron if one neglects isospin violation (and electromagnetic corrections).

\section{\boldmath Behaviour at small $r$}
\label{App:small-r}

Solving the Euler-Lagrange equation (\ref{Eq:diff-eq}) iteratively
at small $r$ one obtains for the profile function the expansion
\be\label{App:profile-small-r}
    F(r) = \sum\limits_{k} \frac{a_k}{k!}\;(e F_\pi r)^k
\ee
where $a_0=\pi$ and $a_k=0$ for even $k\ge 2$. With the definition 
$\widetilde{m}_\pi = m_\pi/(e F_\pi)$ the $a_k$ for odd $k$ are given by
\ba
    a_1 &\equiv & \alpha \phantom{\frac11}
    \nonumber\\
    a_3 &=& -\frac{4\alpha^3(1+2\alpha^2)+3\alpha\widetilde{m}_\pi^2}{
    5(1+8\alpha^2)}
    \nonumber\\
    a_5 &=& \frac{
    24\alpha^5                   (5+32\alpha^2+88\alpha^5+448\alpha^6)
    +3\alpha^3\widetilde{m}_\pi^2(35-16\alpha^2+832\alpha^4)
    +3\alpha  \widetilde{m}_\pi^4(5-56 \alpha^2)}{
    35(1+8\alpha^2)^2}\;,
    \label{App:profile-small-r-2}
\ea
etc. Notice that $\alpha=\alpha(\widetilde{m}_\pi)$.
Numerically we find $\alpha = -1.00376$ for $\widetilde{m}_\pi =0$,
and                 $\alpha = -1.10752$ for $\widetilde{m}_\pi = 0.22715$ 
(which corresponds to $m_\pi =138\,{\rm MeV}$ for our choice of parameters).
In this way  (\ref{Eq:diff-eq}) can be solved in principle
as a differential equation to any finite order in $r$. However, this
iterative procedure can only be used after (\ref{Eq:diff-eq})
is solved as a boundary value problem
(with the constraints $F(0)=\pi$ and $F(r)\to 0$ as $r\to\infty$)
and $\alpha = F^\prime(0)/(e F_\pi)$ is known.

\begin{figure*}[t!]
\begin{tabular}{cccc}
        \includegraphics[height=7cm]{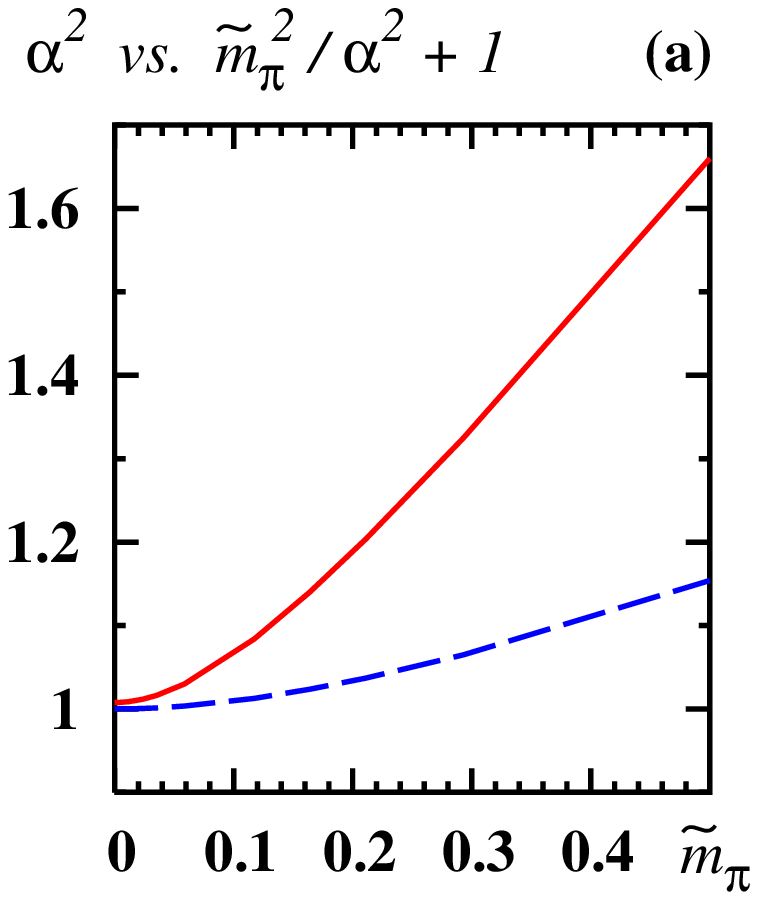} &
    	\includegraphics[height=7cm]{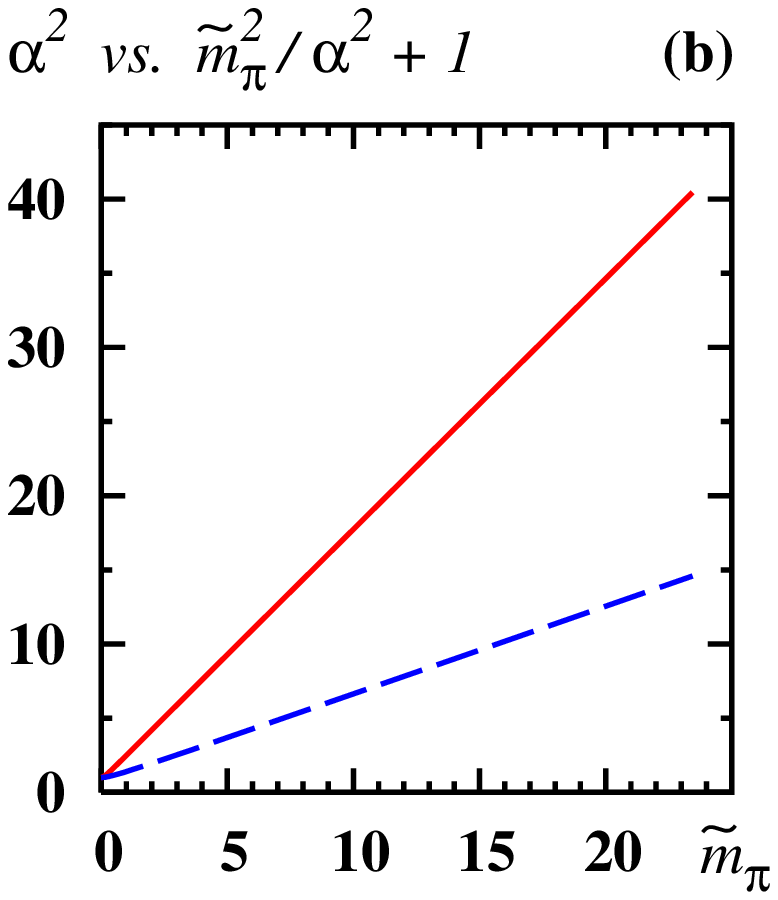}
\end{tabular}
    \caption{\label{Fig10-alpha-mpi}
    \footnotesize\sl
	Solid lines:
	The dimensionless quantity $\alpha = F^\prime(0)/(eF_\pi)$ as
	function of $\widetilde{m}_\pi=m_\pi/(e F_\pi)$ in the region of
	(a) small $\widetilde{m}_\pi\le0.5$, and
	(b) up to $\widetilde{m}_\pi\le 25$.
	Dashed lines:
	The function $\widetilde{m}_\pi^2/\alpha^2+1$ which is smaller than
 	$\alpha^2$ demonstrating that $p(0)>0$, cf.\ Eq.~(\ref{Ineq:p(0)>0}).} 
\end{figure*}

Inserting (\ref{App:profile-small-r},~\ref{App:profile-small-r-2})
into the model expressions (\ref{Eq:T00-Skyrme}-\ref{Eq:shear})
for the densities $T_{00}(r)$, $\rho_J(r)$, $p(r)$ and $s(r)$ one obtains
the results in (\ref{Eq:T00-Skyrme-small-r}-\ref{Eq:rhoJ-Skyrme-small-r})
with the positive constants
\ba
    T_{00}(0) =
    \frac{e^2 F_\pi^4}{8}(3\alpha^2+12\alpha^4+4\widetilde{m}_\pi^2)
    \; ,&& \;\;\;
    A = \frac{e^4 F_\pi^6}{4}\;
    \alpha^2(\alpha^2+4\alpha^4+\widetilde{m}_\pi^2)\\
    \label{App:pressure-small-r-2}
    p(0)    =
    \frac{e^2 F_\pi^4}{8}
    (\;4 \alpha^4-\alpha^2-4 \widetilde{m}_\pi^2)
    \; ,&& \;\;\;
    B=\frac{e^4 F_\pi^6}{4}
    \;\frac{\alpha^2(8\alpha^4-\alpha^2-2\widetilde{m}_\pi^2)(1+ 4\alpha^2)}
    {(1+8\alpha^2)}\\
        && \;\;\;
    C = \frac{e^2 F_\pi^4}{4}\;\frac{\alpha^2(8\alpha^2+1)}{3\Theta} \;.
\ea
From (\ref{App:pressure-small-r-2}) it is not obvious that $p(0)$ and $B$ 
are positive. In order to check that this is really the case, we note first that 
if $p(0)>0$, then also 
$B\propto (8\alpha^4- \alpha^2-2\widetilde{m}_\pi^2) >  
          (8\alpha^4-2\alpha^2-8\widetilde{m}_\pi^2) \propto p(0) >0$.
Thus it is sufficient to demonstrate that $p(0)$ is positive
which can be rewritten as
\be\label{Ineq:p(0)>0}
	p(0) > 0 \;\;\; \Leftrightarrow \;\;\; 
	\alpha^2 > \frac{\widetilde{m}_\pi^2}{\alpha^2}  +  \frac14 \;.
\ee
By numerically solving the boundary value problem for different 
$\widetilde{m}_\pi\le 0.5$, which for our choice of parameters in 
(\ref{Eq:para-fixed}) corresponds to $m_\pi \lesssim 300\,{\rm MeV}$,
we observe that $\alpha^2 > \widetilde{m}_\pi^2/\alpha^2 + 1$, see 
Fig.~\ref{Fig10-alpha-mpi}a. In particular, the weaker inequality 
(\ref{Ineq:p(0)>0}) is satisfied. Thus, for reasonable choices of the
parameters $e$, $F_\pi$, $m_\pi$ we find $p(0) > 0$.

One may wonder whether $p(0)$ is always positive, irrespective the choice 
of $e$, $F_\pi$, $m_\pi$. Recalling that the boundary value problem 
(\ref{Eq:diff-eq}) is characterized by $\widetilde{m}_\pi$ only, 
which follows from substituing $x = r\,e\,F_\pi$ in (\ref{Eq:diff-eq}),
we observe that the inequality (\ref{Ineq:p(0)>0}) is satisfied 
at least up to $\widetilde{m}_\pi \le 25$ (where we start to arrive 
at the limits of our numerical procedure), see Fig.~\ref{Fig10-alpha-mpi}b.
Our observations suggest
that if the boundary value problem (\ref{Eq:diff-eq}) has a solution, 
then it yields $p(0)>0$ (and, as we explicitly checked, 
satisfies (\ref{Eq:stability}), and yields a negative $d_1$). 
It seems that $p(0) > 0$ is a property of the boundary 
value problem (\ref{Eq:diff-eq}), though we are able to verify this
only in a finite range of the (positive) parameter $\widetilde{m}_\pi$
(which, however, goes far beyond what is needed to describe the physical
situation in the Skyrme model).

Another interesting observation is the linear asymptotic behaviour 
$\alpha^2 \approx A\,\widetilde{m}_\pi + B$ with $A\approx 1.688$ 
and $B\approx 0.865$, which sets in already at moderate values 
of $\widetilde{m}$, see Figs.~\ref{Fig10-alpha-mpi}a and b.

\section{\boldmath Chiral properties of the form factors}
\label{App:chiral-properties}

%
\begin{wrapfigure}[17]{R!}{7.2cm}
\vspace{-1cm}
\begin{center}
\includegraphics[width=6cm]{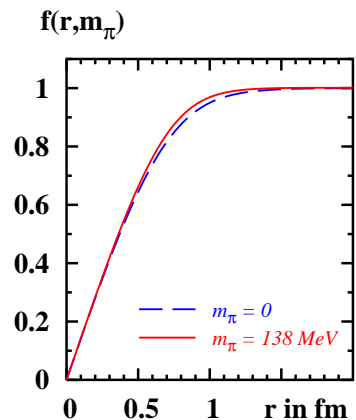}
\end{center}
\vspace{-0.6cm}
	\caption{\label{Fig10-atan-profile}
	\footnotesize\sl
	The function $f(r,m_\pi)$ defined in Eq.~(\ref{Eq:arc-tan-profile}).}
\end{wrapfigure}

In this Appendix we derive the results presented in
Eqs.~(\ref{Eq:chi-prop-d1}-\ref{Eq:chi-prop-Jprime-rJ2}). In particular,
we shall see that these properties are determined by the long distance
behaviour of the profile function (\ref{Eq:prof-large-r}),
i.e.\ that they are of general character.

For that we notice that it is possible to rewrite the profile function as
\be\label{Eq:arc-tan-profile}
    F(r) = 2\,{\rm arctan}\,\biggl[
    \frac{R_0^2}{r^2} \,f(r,m_\pi)\,(1+m_\pi r)\,\exp(-m_\pi r)\biggr]
\ee
with $f(r,m_\pi)$ satisfying the following properties. First, as $r\to\infty$
the function $f(r,m_\pi)\to 1$ while all its derivatives (with respect to $r$
and/or $m_\pi$) tend to zero. This guarantees the correct large distance behaviour
of $F(r)$ in (\ref{Eq:prof-large-r}). Second, the soliton equation of motion
(\ref{Eq:diff-eq}) demands that this function has a Taylor expansion around  $r=0$
of the form $f(r,m_\pi)=\sum_n c_n(m_\pi)\,r^n$ in which only odd powers
$n=1,\,3,\,5,\,\dots$ appear. I.e., it vanishes for small $r$ as
$f(r,m_\pi) \propto r$, which guarantees the correct boundary value for the soliton
profile $F(0)= \pi$ in the Ansatz (\ref{Eq:arc-tan-profile}).

The function $f(r,m_\pi)$ can be determined uniquely from (\ref{Eq:arc-tan-profile})
after Eq.~(\ref{Eq:diff-eq}) has been solved for $F(r)$ with the boundary
values $F(0)=\pi$ and $F(r)\to 0$ as $r\to\infty$. Alternatively, one may
insert the Ansatz (\ref{Eq:arc-tan-profile}) into Eq.~(\ref{Eq:diff-eq})
and solve directly the new differential equation for $f(r,m_\pi)$ subject
to the new boundary conditions: $\lim_{r\to 0} f(r,m_\pi)/r={\rm const}$ and
$f(r,m_\pi)\to 1$ as $r\to\infty$. We stress that by introducing
(\ref{Eq:arc-tan-profile}) we merely change the notation but do not introduce any
approximation. Fig.~\ref{Fig10-atan-profile} shows $f(r,m_\pi)$ for $m_\pi=0$ 
and $138\,{\rm MeV}$.

We discuss first the slope of the form factor $d_1(t)$ at zero momentum transfer.
The general expression for $d_1^\prime(0)$  in terms of $s(r)$ follows from
expanding Eq.~(\ref{Eq:ff-d1}) for small $t$ and reads
\be\label{Eq:d1prime-from-s(r)-and-p(r)}
        d^\prime_1(0)
    = -\,\frac{M_N}{42}\, \int\di^3{\bf r}\;r^4\, s(r)
    =    \frac{M_N}{16}\, \int\di^3{\bf r}\;r^4\, p(r)\;.
\ee
The second relation in (\ref{Eq:d1prime-from-s(r)-and-p(r)}) follows from
integrations by parts and the relation (\ref{Eq:relation-p(r)-s(r)}).
From the large distance asymptotics of $p(r)$ and $s(r)$ in
(\ref{Eq:pressure-large-r},~\ref{Eq:shear-large-r}) it is clear that $d^\prime_1(0)$
diverges in the chiral limit, and is well defined only for $m_\pi\neq 0$ (and that
only for $m_\pi\neq 0$  manipulations like the integration by parts are well defined).

We consider first the expression (\ref{Eq:d1prime-from-s(r)-and-p(r)}) for
$d_1^\prime(0)$ in terms of $s(r)$. For our purposes the model expression
(\ref{Eq:shear}) for the distribution of shear forces can be conveniently
decomposed as  $s(r)= s_A(r)+s_B(r)$ where
\ba\label{Eq:d1prime-00}
    s_A(r) = \frac{F_\pi^2}{4}
    \biggl(F^\prime(r)^2-\frac{\sin^2F(r)}{r^2}\biggr)\; , \;\;\;
    s_B(r) = \frac{\sin^2F(r)}{e^2\,r^2}
    \biggl(F^\prime(r)^2-\frac{\sin^2F(r)}{r^2}\biggr)\;.
\ea
In the chiral limit $s_B(r)\propto \frac{1}{r^{12}}$ at large $r$, and gives
a contribution to the slope of $d_1(t)$ at $t=0$ which is well defined and
finite in the chiral limit. Thus, denoting this contribution by
$d_1^\prime(0)_B$, we note that
\be\label{XXEq:d1prime-01}
    d_1^\prime(0)_B = -\frac{M_N}{42}\int\limits_0^\infty
    \di^3{\bf r}\;r^4 \,s_B(r) = {\mathcal O}(m_\pi^0).
\ee
In fact, what makes $d_1^\prime(0)$ divergent in the chiral limit it is the part
$s_A(r)\propto \frac{1}{r^6}$ at large $r$. Let $m_\pi\neq 0$ in the following.
We perform the integral over $r$ in a spherical box of radius $D$,
and take $D\to\infty$ later.
Inserting the expression (\ref{Eq:d1prime-00}) for $s_A(r)$ with the Ansatz
(\ref{Eq:arc-tan-profile}) into (\ref{Eq:d1prime-from-s(r)-and-p(r)}) and making
the legitimate (because $m_\pi\neq 0$) substitution $r\to a/m_\pi$ yields
\be\label{Eq:d1prime-02}
    d_1^\prime(0)_A =
    \frac{4\pi M_N R_0^4 F_\pi^2}{42 m_\pi}
    \int\limits_0^{D m_\pi} \!\!\!\di a\; G_1(a,m_\pi)
\ee
where we defined the function (with
$f^\prime(y,m_\pi)\equiv \frac{\partial \;}{\partial y}f(y,m_\pi)$ for brevity)
\ba\label{Eq:d1prime-03}
    G_1(a,m_\pi) &=&
    \frac{a^8 e^{-2a}}
    {\left(a^4+ m_\pi^4 R_0^4 \,(1+a)^2 f(\frac{a}{m_\pi}, m_\pi )^2 e^{-2a}
    \right)^2}
    \times\biggl\{
    \left(a^4+4a^3+7a^2+6 a+3\right)f(\fracS{a}{m_\pi},m_\pi)^2 \nonumber\\
&&      -2a\left(a^3+3a^2+4a+2\right)
    f(\fracS{a}{m_\pi},m_\pi)f^\prime(\fracS{a}{m_\pi},m_\pi)
    +a^2(a+1)^2 f^\prime(\fracS{a}{m_\pi},m_\pi)^2 \biggr\}\;.
\ea
In Eq.~(\ref{Eq:d1prime-02}) for $m_\pi\neq 0$ we may safely take $D\to\infty$.
This step is trivial in our analytical calculation, however,
it is a subtle issue for calculations (in models or lattice QCD) performed
in a finite volume. The correct chiral behaviour is obtained only by first taking the
infinite volume limit, and considering then the limit of pion masses becoming small.
The $m_\pi$-expansion of the integral (\ref{Eq:d1prime-02}) reads
\be\label{Eq:d1prime-04}
    \int\limits_0^\infty \!\di a\; G_1(a,m_\pi)
    = \int\limits_0^\infty \!\di a\; e^{-2a}
    \left(a^4+4a^3+7a^2+6 a+3\right) + {\mathcal O}(m_\pi)
    = 7  + {\mathcal O}(m_\pi) \;,
\ee
where for the leading term we made use of the fact that $f(y,m_\pi)\to 1$
for large arguments $y$, while its derivatives go to zero. Eliminating $R_0$ in
favour of $g_A$ and $F_\pi$ according to (\ref{Eq:prof-large-r}) we obtain
the result quoted in (\ref{Eq:chi-prop-d1prime}).
We may repeat this exercise exploring the relation of $d_1^\prime(0)$ in terms
of the pressure in Eq.~(\ref{Eq:d1prime-from-s(r)-and-p(r)}).
The calculation is analog to that presented above and yields the same result
(\ref{Eq:chi-prop-d1prime}).
Thus, our procedure to derive the leading non-analytic chiral
contribution to $d_1^\prime(0)$ {\sl respects} the model equations of motion.

Next let us discuss the chiral expansion of $d_1 \equiv d_1(t)|_{t=0}$.
Since $s(r)$ and $p(r)\propto \frac{1}{r^6}$ at large $r$, see
(\ref{Eq:pressure-large-r},~\ref{Eq:shear-large-r}),
$d_1$ takes a finite value in the chiral limit.
The next term in its chiral expansion is linear in $m_\pi$
which can be found as follows.
We use the relation (\ref{Eq:d1-from-s(r)-and-p(r)}) for $d_1$ in terms of
$s(r)$ and the presentation (\ref{Eq:arc-tan-profile}) for the soliton profile,
and interchange the order of differentiating with respect to $m_\pi$ and
integrating over $r$.  Substituting $r\to a/m_\pi$ we obtain
\be\label{Eq:d1-mpi-01}
    \frac{\partial \, d_1}{\partial m_\pi}
    = -\,\frac{1}{3}\int\di^3{\bf r}\;r^2 \frac{\partial \,s(r)}{\partial m_\pi}
    = -\,\frac{4\pi M_N F_\pi^2R_0^4}{3}
    \int\limits_0^\infty \!\di a\; G_2(a,m_\pi) \;,
\ee
where $G_2(a,m_\pi)$ is a function similar to (but much lengthier than) that
in Eq.~(\ref{Eq:d1prime-03}), whose small $m_\pi$-expansion reads
\be\label{XXEq:d1-mpi-02}
    \int\limits_0^\infty \!\di a\; G_2(a,m_\pi)
    = -2 \int\limits_0^\infty \!\di a\; e^{-2a}
    \left(a^3+2a^2+a-1\right) + {\mathcal O}(m_\pi)
    = -\frac54  + {\mathcal O}(m_\pi) \;,
\ee
which yields the chiral expansion of $d_1$ quoted in Eq.~(\ref{Eq:chi-prop-d1}).
Starting from the expression (\ref{Eq:d1-from-s(r)-and-p(r)}) which expresses
$d_1$ in terms of $p(r)$ one reproduces the result in Eq.~(\ref{Eq:chi-prop-d1})
by a similar calculation. We stress that also the calculations leading to
(\ref{Eq:chi-prop-d1}) respect the model equations of motion.

Next we consider the mean square radius of the energy density
defined in (\ref{Eq:def-energy-mean-square-radius}).
Since $T_{00}(r)\propto\frac{1}{r^6}$ at large $r$, see
Eq.~(\ref{Eq:T00-Skyrme-large-r}), the mean square radius of the energy density
has a well defined chiral limit. Its first chiral correction appears already
in linear order in $m_\pi$ and the coefficient of this leading non-analytic term
can be found  by means of a calculation which is completely analog to that in
Eq.~(\ref{Eq:d1-mpi-01}) and yields the result quoted in Eq.~(\ref{Eq:chi-prop-rE2}).

Finally, we consider the mean square radius of the angular momentum density
which is defined in (\ref{Eq:def-ang-mom-mean-square-radius}).
A calculation analog to that in Eqs.~(\ref{Eq:d1prime-00}-\ref{Eq:d1prime-04})
yields the result quoted in Eq.~(\ref{Eq:chi-prop-Jprime-rJ2}).

\section{Scaling relations}
\label{App:scaling-relations}

In our approach --- in which we consider $1/N_c$ corrections to some quantity
if and only if the respective leading order contribution vanishes, 
see Sec.~\ref{Sec-3:model} --- the dependence on the model parameters
$F_\pi$ and $e$ is trivial and given by the following scaling relations
\ba
M_N 	    =\frac{F_\pi}{e}     \;G\biggl(\frac{m_\pi}{e F_\pi}\biggr),&\;\;\;\;\;\;
\displaystyle
\Theta      =\frac{e^3}{F_\pi}   \;G\biggl(\frac{m_\pi}{e F_\pi}\biggr),&\;\;\;\;\;\;
g_A 	    =\frac{1}{e^2}       \;G\biggl(\frac{m_\pi}{e F_\pi}\biggr),   \nonumber\\
\mu_{I=0}   =                      G\biggl(\frac{m_\pi}{e F_\pi}\biggr),&\;\;
\displaystyle
\mu_{I=1}   =\frac{1}{e^4}       \;G\biggl(\frac{m_\pi}{e F_\pi}\biggr),&\;\;\;\;\;\;\;
d_1 	    =\frac{1}{e^4}       \;G\biggl(\frac{m_\pi}{e F_\pi}\biggr),
							\label{Eq-scaling-relations}\\
T_{00}(0)   =\frac{F_\pi^4}{e^2} \;G\biggl(\frac{m_\pi}{e F_\pi}\biggr),&\;\;
\displaystyle
p(0) 	    =\frac{F_\pi^4}{e^2} \;G\biggl(\frac{m_\pi}{e F_\pi}\biggr),&
\la r_i^2\ra=\frac{1}{F_\pi^2e^2}\;G\biggl(\frac{m_\pi}{e F_\pi}\biggr).    \nonumber
\ea
The last result holds for all mean square radii.
The functions $G(y)$ are different for the different quantities.

If one chooses $e$, $F_\pi$ differently from the parameter fixing 
used in this work in Eqs.~(\ref{Eq:para-fixing},~\ref{Eq:para-fixed}), 
then the relations (\ref{Eq-scaling-relations}) allow to rescale 
the results for the respective quantities in terms of the new parameters.
Strictly speaking one needs for that the $m_\pi$-dependence
of the respective quantities which we have provided in 
Sec.~\ref{Sec-7:Results-ffs} for quantities of interest.


\end{document}